\documentclass[
 pra,twocolumn,
 amsmath,amssymb,
 aps,nofootinbib,longbibliography,
]{revtex4-2}

\usepackage{amsthm}
\usepackage{graphicx}
\usepackage{color}
\usepackage{dcolumn}
\usepackage{bm}
\usepackage{mathtools}      
\usepackage{mathrsfs}       
\usepackage{braket}
\usepackage{float}
\usepackage[table]{xcolor}

\newcommand{\cA}{{\mathcal{A}}}

\newcommand{\cD}{{\mathcal{D}}}
\newcommand{\cE}{{\mathcal{E}}}

\newcommand{\cH}{{\mathcal{H}}}
\newcommand{\cN}{{\mathcal{N}}}
\newcommand{\cP}{{\mathcal{P}}}
\newcommand{\cQ}{{\mathcal{Q}}}
\newcommand{\cR}{{\mathcal{R}}}

\newcommand{\cK}{{\mathcal{K}}}
\newcommand{\cL}{{\mathcal{L}}}

\newcommand{\cM}{{\mathcal{M}}}
\newcommand{\cX}{{\mathcal{X}}}
\newcommand{\cY}{{\mathcal{Y}}}

\newcommand{\cT}{{\mathcal{T}}}

\newcommand{\bx}{{\boldsymbol{x}}}
\newcommand{\by}{{\boldsymbol{y}}}

\newcommand{\bh}{{\boldsymbol{h}}}

\newcommand{\br}{{\boldsymbol{r}}}

\newcommand{\baph}{{\boldsymbol{\alpha}}}
\newcommand{\bbta}{{\boldsymbol{\beta}}}
\newcommand{\bphi}{{\boldsymbol{\phi}}}

\newcommand{\bth}{{\boldsymbol{\theta}}}

\newcommand{\bH}{{\boldsymbol{H}}}

\newcommand{\bb}{{\boldsymbol{b}}}

\newcommand{\bw}{{\boldsymbol{w}}}

\newcommand{\bbR}{{\mathbb{R}}}
\newcommand{\bbE}{{\mathbb{E}}}

\newcommand{\hrba}{\hat{r}_{\boldsymbol{\alpha}}}

\DeclareMathOperator{\tr}{\textup{Tr}}

\DeclareMathAlphabet{\mymathbb}{U}{BOONDOX-ds}{m}{n}
\usepackage[utf8]{inputenc} 

\usepackage{hyperref}
\hypersetup{
    colorlinks = true,
    linkcolor = {blue},
    citecolor = {blue},
    urlcolor = {blue},
}

\begin{document}
\title{Quantum Curriculum Learning}

\author{Quoc Hoan Tran}
\email{tran.quochoan@fujitsu.com}
\affiliation{Quantum Laboratory, Fujitsu Research, Fujitsu Limited, Kawasaki, Kanagawa 211-8588, Japan}

\author{Yasuhiro Endo}
\affiliation{Quantum Laboratory, Fujitsu Research, Fujitsu Limited, Kawasaki, Kanagawa 211-8588, Japan}

\author{Hirotaka Oshima}
\affiliation{Quantum Laboratory, Fujitsu Research, Fujitsu Limited, Kawasaki, Kanagawa 211-8588, Japan}

\date{\today}

\begin{abstract}

Quantum machine learning (QML) requires significant quantum resources to address practical real-world problems. When the underlying quantum information exhibits hierarchical structures in the data, limitations persist in training complexity and generalization. Research should prioritize both the efficient design of quantum architectures and the development of learning strategies to optimize resource usage.
We propose a framework called quantum curriculum learning (Q-CurL) for quantum data, where the curriculum introduces simpler tasks or data to the learning model before progressing to more challenging ones. Q-CurL exhibits robustness to noise and data limitations, which is particularly relevant for current and near-term noisy intermediate-scale quantum devices. We achieve this through a curriculum design based on quantum data density ratios and a dynamic learning schedule that prioritizes the most informative quantum data.
Empirical evidence shows that Q-CurL significantly enhances training convergence and generalization for unitary learning and improves the robustness of quantum phase recognition tasks. Q-CurL is effective with physical learning applications in physics and quantum chemistry. 
\end{abstract}

\pacs{Valid PACS appear here}

\maketitle

\section{Introduction} 
In the emerging field of quantum computing (QC), there is potential to use large-scale quantum computers to solve certain machine learning (ML) problems far more efficiently than classical methods. This synergy between ML and QC has given rise to quantum machine learning (QML)~\cite{biamonte:2017:QML,schuld:2021:qmlbook}, although its practical applications remain uncertain.
Early QML research focused on quantum algorithms that theoretically enhance the efficiency of linear algebra subroutines critical to ML. A notable example is the Harrow-Hassidim-Lloyd (HHL) algorithm~\cite{HHL:2009:prl}, which is designed to solve large systems of linear equations exponentially faster than classical computers. However, the HHL algorithm's potential relies on careful preconditioning~\cite{scott:2015:natphys} to accelerate quantum computations on qubits, without considering the time required for input/output processes. These processes involve loading classical data into quantum states and extracting classical solutions from quantum states, which can be prohibitively slow, potentially negating the quantum speedup. Furthermore, if classical algorithms can efficiently utilize computational basis measurements required by a quantum algorithm, they can also exploit these measurements to accelerate linear algebra operations, rendering computation time independent of the problem's dimensionality. This concept, known as dequantization~\cite{tang:2022:natphys}, underscores a significant challenge to achieving quantum advantage.

Classical ML traditionally focuses on extracting and replicating features based on data statistics.
Inspired by the success of deep learning, QML with the central techniques like quantum circuit learning~\cite{mitarai:2018:circuit,schuld:2019:shift} with variational quantum algorithms~\cite{cerezo:2021:variational:nature}, quantum kernel with quantum feature maps~\cite{halvlicek:2019:supervised, schuld:2019:feature}, and quantum reservoir computing~\cite{fujii:2017:qrc} with input-driven quantum dynamics, has drawn much attention in recent years.
In these approaches, the central concept is to transform the task of learning from classical data into the task of identifying distinguishing features within quantum states in Hilbert space.
QML is hoped to detect correlations in classical data or generate patterns that are challenging for classical algorithms to achieve~\cite{halvlicek:2019:supervised,schuld:2019:feature,liu:2020:rigorous,tran:2021:prl:uap,gao:2022:prx:corr,tran:2020:higherorder}. However, it remains unclear whether analyzing classical data fundamentally requires quantum effects.
Furthermore, there is a question as to whether speed is the only metric by which QML algorithms should be judged~\cite{schuld:2022:qadv}. 
This suggests a fundamental shift: it is preferable to use QML on data that is already quantum in nature~\cite{editorial:2023:qmladv:nature,cong:2019:QCNN,Perrier:2022:qdata,haug:2023:generalization,chinzei:2024:spQCNN,tran:2024:varQAE}. 

\begin{figure*}
		\includegraphics[width=18cm]{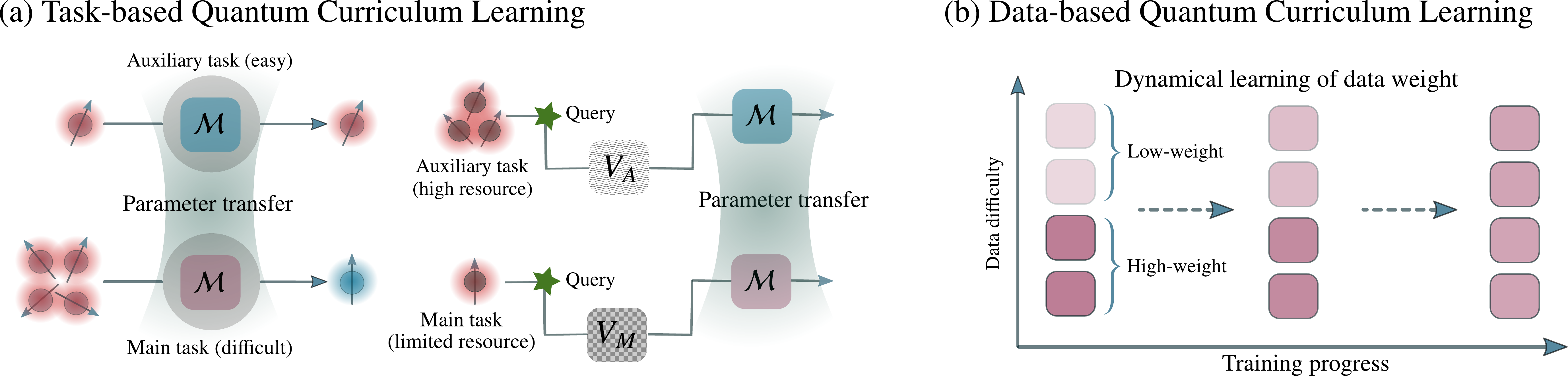}
		\protect\caption{Overview of two principal methodologies in quantum curriculum learning: (a) task-based and (b) data-based approaches. In the task-based approach, a model $\cM$, designated for a main task that may be challenging or constrained by data accessibility, benefits from pre-training on an auxiliary task. This auxiliary task is either relatively simpler (left panel of (a)) or has a richer dataset (right panel of (a)).
        In the task-based approach, parameter transfer is executed by initializing the model parameters of the main task with the optimal parameters derived from the auxiliary task.
        In the data-based approach, we implement a dynamic learning schedule to modulate data weights, thereby emphasizing the significance of quantum data in optimizing the loss function to reduce the generalization error.
		\label{fig:qcurr:overview}}
\end{figure*}

The learning process in QML involves extensive exploration within the domain landscape of a loss function. This function measures the discrepancy between the quantum model's predictions and the actual values, aiming to locate its minimum. However, the optimization often encounters pitfalls such as getting trapped in local minima~\cite{bittel:2021:prl:VQANP,ansachuetz:2022:natcom:VQA} or barren plateau regions~\cite{clean:2018:natcom:barren}. These scenarios require substantial quantum resources to navigate the loss landscape successfully. Additionally, improving accuracies necessitates evaluating numerous model configurations, especially against extensive datasets. Given the limitation of quantum resources in designing QML models, we must focus not only on their architectural aspects but also on efficient learning strategies.

The perspective of quantum resources refocuses our attention on the concept of learning. In ML, learning refers to the process through which a computer system enhances its performance on a specific task over time by acquiring and integrating knowledge or patterns from data. We can improve current QML algorithms by making this process more efficient.
This approach aligns with the intriguing perspective that describes intelligence as the efficiency of skill acquisition~\cite{chollet:2019:intelligence}.
Curriculum learning, inspired by human educational strategies, offers a promising framework to achieve such efficiency. Proposed by Bengio et al.~\cite{bengio:2009:curr}, curriculum learning involves presenting training data or tasks in a structured order, progressing from simpler to more complex examples, thereby facilitating more effective learning. This strategy mirrors human learning, where foundational concepts are mastered before tackling advanced ones, enabling a model to build robust representations incrementally.

In classical ML, various types of curriculum learning has been widely adopted to improve training efficiency and model performance across various domains~\cite{soviany:2022:curr:survey}. 
Vanilla curriculum learning, introduced by Bengio et al.~\cite{bengio:2009:curr}, utilizes rule-based criteria to order samples from simple to complex, thereby improving the convergence of training. Balanced curriculum learning adds diversity constraints to ensure varied samples at each stage, preventing overfitting~\cite{Zhang:2015:diversity:curr,soviany:2020:curriculum}. Self-paced curriculum learning, introduced by Jiang et al.~\cite{jiang:2015:self-paced}, combines predefined and learning-based criteria for tasks like matrix factorization and multimedia event detection. Progressive curriculum learning applies curriculum concepts to model capacity or task settings, as seen in Karras et al.~\cite{karras:2018:progressive} for growing Generative Adversarial Networks. Teacher-student curriculum learning, proposed by Matiisen et al.~\cite{matiisen:2020:teacher}, uses an auxiliary model to guide the primary model's learning parameters, optimizing training policy. Implicit curriculum learning, exemplified by Sinha et al.~\cite{sinha:2020:implicit}, integrates curriculum effects indirectly, such as gradually deblurring activation maps, enhancing complexity without explicit sample ordering.
Techniques such as data parameters~\cite{saxena:2019:dataparm:NIPS}, loss-based sample weighting~\cite{superloss:2020:NIPS} and dynamic curriculum scheduling~\cite{wang:2019:dynamic:curr} further enhance training by adaptively prioritizing samples based on their difficulty or relevance. These approaches have been successfully applied in computer vision to prioritize simpler images or features during training~\cite{novot:2018:CVPR}, in natural language processing to sequence samples by difficulty~\cite{xu:metal:2020:curriculum}, and in reinforcement learning to guide agents through progressively challenging environments~\cite{narvekar:2020:currsurvey}.


Although curriculum learning has been extensively applied in classical ML, its exploration in the QML field, especially regarding quantum data, is still in the early stages. Existing research has primarily examined model transfer learning in hybrid classical-quantum networks~\cite{mari2020:quantum:transfer}, where a pre-trained classical model is enhanced by adding a variational quantum circuit. 
During the revision of this manuscript, we identified a relevant application of curriculum learning in quantum circuit architecture search for determining the ground state of a given Hamiltonian~\cite{kundu:2025:tensorrlqas}. This approach employs a warm-start strategy, initializing training with an easily prepared approximate state before refining the solution using reinforcement learning.
However, there is still limited evidence showing that curriculum learning can effectively improve QML by scheduling tasks and samples.

We explore the potential of curriculum learning using quantum data. 
We implement a quantum curriculum learning (Q-CurL) framework in two common scenarios. First, a main quantum task, which may be challenging due to the high-dimensional nature of the parameter space or the limitation of data availability, can be facilitated through the hierarchical parameter adjustment of auxiliary tasks. These auxiliary tasks are comparatively easier or more data-rich.
Here, the parameters learned from the auxiliary task serve as an initial configuration for the parameters of the main task. This approach is particularly significant in the context of quantum data learning, where preparing perfect, noiseless quantum states is challenging. By leveraging noisy quantum data or easily prepared quantum states in auxiliary tasks, the learning process for the main task can be accelerated, even with a limited amount of training data.
However, it is necessary to establish the criteria that make an auxiliary task beneficial for a main task.
Second, QML often involves noisy inputs that exhibit a hierarchical arrangement of entanglement or noisy labels, reflecting levels of importance during the optimization process. Recognizing these levels is essential for ensuring the robustness and reliability of QML methods in practical scenarios.

We propose two principal approaches to address the outlined scenarios: task-based Q-CurL [Fig.~\ref{fig:qcurr:overview}(a)] for the first and data-based Q-CurL [Fig.~\ref{fig:qcurr:overview}(b)] for the second scenario.
In task-based Q-CurL, the curriculum order is defined by the quantum-based kernel density ratio between quantum datasets. This enables efficient auxiliary task selection without solving each one, reducing data demands for the main task and decreasing training epochs, even if total data requirements stay constant.
In data-based Q-CurL, we employ a dynamic learning schedule that adjusts data weights to prioritize quantum data in optimization. This adaptive cost function is broadly applicable to any cost function without requiring additional quantum resources.
Empirical evidence shows that task-based Q-CurL enhances training convergence and generalization when learning complex unitary dynamics. Additionally, data-based Q-CurL increases robustness, particularly in noisy-label scenarios, by preventing complete memorization of the training data. This avoids overfitting and improves generalization in the quantum phase detection task. These results suggest that Q-CurL could be broadly effective for physical learning applications.

\section{Method}
\subsection{Task-based Q-CurL}
We formulate a framework for task-based Q-CurL. 
In classical ML, it is well-known that learning from multiple tasks can lead to better and more efficient algorithms. This idea encompasses areas such as transfer learning, multitask learning, and meta-learning, all of which have significantly advanced deep learning.
Unlike classical ML, which typically assumes a fixed amount of training data for all tasks, in quantum learning, the order of tasks and the allocation of training data to each task are even more critical. Properly scheduling tasks could reduce the resources required for training the main task, bringing QML closer to practical applications.

The target of learning is to find a function (or hypothesis) $h:\cX \to \cY$ within a hypothesis set $\cH$ that approximates the true function $f$ mapping $\bx \in \cX$ to $\by=f(\bx) \in \cY$. To evaluate the correctness of $h$ given the data $(\bx, \by)$, the loss function $\ell:\cY \times \cY \to \mathbb{R}$ is used to measure the approximation error $\ell(h(\bx), \by)$ between the prediction $h(\bx)$ and the target $\by$.
We aim to find $h \in \cH$ that minimizes the expected risk over the data generation distribution $P(\cX, \cY)$:
\begin{align}\label{eqn:risk}
R(h) := \bbE_{(\bx,\by)\sim P(\cX, \cY)} \left[ \ell(h(\bx), \by) \right].
\end{align}
In practice, since $P(\cX, \cY)$ is unknown, we use the observed dataset $\cD={(\bx_i, \by_i)}_{i=1}^N \subset \cX \times \cY$ to minimize the empirical risk, defined as the average loss over the training data:
\begin{align}\label{eqn:emp:risk}
\hat{R}(h)  = \frac{1}{N}\sum_{i=1}^N\ell(h(\bx_i), \by_i) = \frac{1}{N}\sum_{i=1}^N\ell_i,
\end{align}
where $\ell_i = \ell(h(\bx_i), \by_i)$ is the single loss corresponding with the training data $(\bx_i, \by_i)$.

In our study, consistent with traditional approaches, the hypothesis set $\cH$ is defined as as a parametric collection of hypotheses, denoted as as $\cH = \{h_{\bth}\}$, where $\bth$ represents the model parameters, and each value of $\bth$ corresponds to a specific hypothesis $h_{\bth}$. The objective is to initialize the model parameters at an appropriate starting point, $\bth=\bth_0$, and iteratively optimize them to identify the optimal parameters, $\bth = \bth_{\textup{opt}}$, that minimize the empirical risk $\hat{R}(h_\bth)$.

\subsubsection{Design a curriculum via curriculum weights}

Given a main task $\cT_{M}$, the goal of task-based Q-CurL is to design a curriculum for solving auxiliary tasks to enhance performance compared to solving the main task alone. 
We consider $\cT_1, \ldots, \cT_{M-1}$ as the set of auxiliary tasks. The training dataset for  $\cT_m$ is $\cD_m \subset \cX^{(m)} \times \cY^{(m)}$ ($m=1, \ldots, M$), containing $N_m$ data pairs. We focus on supervised learning tasks with input quantum data $\bx^{(m)}_i$ in the input space $\cX^{(m)}$ and corresponding target data $\by^{(m)}_i$ in the output space $\cY^{(m)}$ for $i=1, \ldots, N_m$.
The training data $\left(\bx^{(m)}_i, \by^{(m)}_i\right)$ for $\cT_m$ are drawn from the probability distribution $P^{(m)}(\cX^{(m)}, \cY^{(m)})$ with the density $p^{(m)}(\cX^{(m)}, \cY^{(m)})$. We assume that all tasks share the same data spaces $\cX^{(m)} \equiv \cX$ and $\cY^{(m)} \equiv \cY$, as well as the same hypothesis class $\{h_{\bth}\}$ and the same loss function $\ell$ for all $m$.

We focus on identifying an auxiliary task $\cT_m$ such that solving $\cT_m$ facilitates the solution of the main task $\cT_M$. For clarity, we denote the parameters of task $\cT_m$ as $\bth^{(m)}$. In this curriculum scheme, only model parameters are transferred. Specifically, upon solving $\cT_m$, we obtain the optimal parameter $\bth^{(m)} = \bth^{(m)}_{\textup{opt}}$.
These parameters are then used as the initial values for the main task, setting $\bth^{(M)}=\bth^{(M)}_0=\bth^{(m)}_{\textup{opt}}$, before iteratively optimizing to achieve the optimal parameters $\bth^{(M)} = \bth^{(M)}_{\textup{opt}}$ for $\cT_M$.

Depending on the problem, we can decide the \textit{curriculum weight} $c_{M,m}$, where a larger $c_{M,m}$ indicates a greater benefit of solving $\cT_m$ for improving the performance on $\cT_M$. We evaluate the contribution of solving task $\cT_i$ to the main task $\cT_M$ by transforming the expected risk of training $\cT_M$ as follows:
\begin{align}
    R_{T_M}(h) &= \bbE_{(\bx,\by)\sim P^{(M)}}\left[\ell(h(\bx), \by) \right]\\
    &= \int\int_{(\bx, \by)}\ell(h(\bx), \by)p^{(M)}(\bx, \by)d(\bx, \by)\\
    &= \int\int_{(\bx, \by)}\dfrac{p^{(M)}(\bx, \by)}{p^{(m)}(\bx, \by)}\ell(h(\bx), \by)p^{(m)}(\bx, \by)d(\bx, \by)\\
    &= \bbE_{(\bx,\by)\sim P^{(m)}}\left[\dfrac{p^{(M)}(\bx, \by)}{p^{(m)}(\bx, \by)}\ell(h(\bx), \by) \right].
\end{align}

The curriculum weight $c_{M,m}$ can be determined using the density ratio $r(\bx, \by) = \dfrac{p^{(M)}(\bx, \by)}{p^{(m)}(\bx, \by)}$ without requiring the density estimation of $p^{(M)}(\bx, \by)$ and $p^{(m)}(\bx, \by)$. Similar to the unconstrained least-squares importance fitting approach~\cite{kanamori:2009:jmlr} in classical ML, the key idea is to model the density ratio function $r(\bx, \by)$ using a linear model:
\begin{align}
\hat{r}_{\baph}(\bx, \by) := \baph^\top\bphi(\bx, \by) = \sum_{i=1}^{N_M} \alpha_i \phi_i(\bx, \by),
\end{align}
where the vector of basis functions is $\bphi(\bx, \by) = (\phi_1(\bx, \by), \ldots, \phi_{N_M}(\bx, \by))$, and the parameter vector $\baph = (\alpha_1, \ldots, \alpha_{N_M})^\top$ is learned from data.

The basis function $\phi_l(\bx, \by)$ is defined as the product of kernels used to compare two pairs of input and output states as:
\begin{align}
\phi_l(\bx, \by) = \cK_x[\bx \bx^{(M)}_l]  \cK_y[\by \by^{(M)}_l].
\end{align}
Here, $\cK_x(\cdot,\cdot)$ and $\cK_y(\cdot, \cdot)$ are the kernels defined in the data space $\cX$ and $\cY$. The key factor that differentiates this framework from classical curriculum learning is the consideration of quantum data for $\bx$ and $\by$, which are assumed to be in the form of density matrices representing quantum states. For example, the kernel function $\cK_x$ and $\cK_y$ can be naturally defined as the global fidelity kernel, which leads to the form of $\phi_l(\bx, \by)$ as
\begin{align}\label{eqn:basis:trace}
\phi_l(\bx, \by) = \tr[\bx \bx^{(M)}_l]  \tr[\by \by^{(M)}_l].
\end{align}

In our numerical experiments, which focus on learning unitary dynamics using quantum data for both input and output, fidelity emerges as an appropriate metric due to its direct relevance to the task. However, to enhance scalability for large-scale quantum data, efficient quantum kernels such as the quantum projected kernel~\cite{huang:2021:power} or the shadow tomography kernel~\cite{huang:2022:science} can be utilized to estimate the density ratio effectively. 
Moreover, in other ML scenarios such as classifying quantum data with quantum inputs and classical outputs, a hybrid approach may prove more effective. In these cases, a quantum kernel can be applied to the quantum components, while a classical kernel complements the classical parts.

In this way, $R_{T_M}(h)$ can be approximated by 
\begin{align}\label{eqn:risk:transfer}
    R_{T_M}(h) \approx \bbE_{(\bx,\by)\sim P^{(m)}}\left[\hrba(\bx, \by)\ell(h(\bx), \by) \right],
\end{align}
or, as an approximation, using the following sample averages:
\begin{align}\label{eqn:transfer:min}
    R_{T_M}(h) \approx \dfrac{1}{N_m}\sum_{i=1}^{N_m} \hrba(\bx^{(m)}_i, \by^{(m)}_i)\ell(h(\bx^{(m)}_i), \by^{(m)}_i).
\end{align}

The parameter vector $\baph$ is estimated by minimizing the following error:
\begin{align}
    &\dfrac{1}{2}\int\int\left[\hrba(\bx, \by) - r(\bx, \by) \right]^2 p^{(m)}(\bx, \by)d\bx d\by\\
    \quad &=\dfrac{1}{2}\int\int \hrba(\bx, \by)^2p^{(m)}(\bx, \by)d\bx d\by - \nonumber \\ 
    &\int \hrba(\bx, \by)p^{(M)}(\bx, \by)d\bx d\by + C \label{dens:est}.
\end{align}
Given the training data, we can further reduce the minimization of Eq.~\eqref{dens:est} to the  problem of minimizing
\begin{align}\label{eqn:min}
    \dfrac{1}{2N_m}\sum_{i=1}^{N_m} \hrba^2(\bx^{(m)}_i, \by^{(m)}_i) - \dfrac{1}{N_M}\sum_{i=1}^{N_M} \hrba(\bx^{(M)}_i, \by^{(M)}_i) + \dfrac{\lambda}{2}\| \baph \|_2^2,
\end{align}
where we consider the regularization coefficient $\lambda$ for $L_2$-norm of $\baph$.
Equation~\eqref{eqn:min} can be further reduced to the following quadratic form:
\begin{align}
    \min_{\baph}\dfrac{1}{2}\baph^\top\bH\baph - \bh^\top\baph + \dfrac{\lambda}{2}\baph^\top\baph.
\end{align}
Here, $\bH$ is the $N_M\times N_M$ matrix with elements $H_{ll^\prime}=\dfrac{1}{N_m}\sum_{i=1}^{N_m}\phi_l(\bx_i^{(m)}, \by_i^{(m)})\phi_{l^\prime}(\bx_i^{(m)}, \by_i^{(m)})$,
and $\bh$ is the $N_M$-dimensional vector with elements $h_l = \frac{1}{N_M}\sum_{i=1}^{N_M}\phi_l(\bx_i^{(M)}, \by_i^{(M)})$.

We can consider each $\hat{r}(\bx^{(m)}_i, \by^{(m)}_i)$ in Eq.~\eqref{eqn:transfer:min} as the contribution of the data $(\bx^{(m)}_i, \by^{(m)}_i)$ from the auxiliary task $\cT_m$ to the main task $\cT_M$. 
From Eq.~\eqref{eqn:transfer:min}, we note that only the quantity $\ell(h(\bx^{(m)}_i), \by^{(m)}_i)$ depends on the training performance of the auxiliary task $\cT_m$. We assume that the loss $\ell(h(\bx^{(m)}_i), \by^{(m)}_i)$ is bounded by a quantity $\ell_{\textup{max}}^{(m)}$ for all $i=1,\ldots,N_m$. Then the empirical risk $R_{T_M}(h)$ (before solving $\cT_M$) can be bounded by the following inequality:
\begin{align}
    R_{T_M}(h) &\approx \dfrac{1}{N_m}\sum_{i=1}^{N_m} \hrba(\bx^{(m)}_i, \by^{(m)}_i)\ell(h(\bx^{(m)}_i), \by^{(m)}_i) \\
    &\leq \dfrac{\ell_{\textup{max}}^{(m)}}{N_m}\sum_{i=1}^{N_m} \hrba(\bx^{(m)}_i, \by^{(m)}_i) = \ell^{(m)}_{\textup{max}} c_{M,m}\label{eqn:risk:bound},
\end{align}
where the curriculum weight $c_{M,m}$ is defined as
\begin{align}\label{eqn:curr:weight}
    c_{M,m} = \dfrac{1}{N_m}\sum_{i=1}^{N_m} \hrba(\bx^{(m)}_i, \by^{(m)}_i).
\end{align}

We clarify that $c_{M,m}$ quantifies the effect of minimizing $\ell^{(m)}_{\textup{max}}$, associated with the auxiliary task $\cT_m$, on the empirical risk in training the main task $\cT_M$.
In Eq.~(17), the empirical risk is bounded with the product of $l^{(m)}_{\textup{max}}$ and $c_{M,m}$.
This upper bound is independent of the optimization process for the main task $\cT_M$.
Here, $\ell^{(m)}_{\textup{max}}$ is determined by the optimization process of the auxiliary task $\cT_m$, while $c_{M,m}$ reflects the similarity between the data domains of the auxiliary and main tasks.
We compare the upper bound in two scenarios: one without solving $\cT_m$ in advance, 
and another where $\cT_m$ is solved first, with its parameters transferred to $\cT_M$.
We focus on assessing the extent to which solving $\mathcal{T}_m$ reduces this upper bound better.
More concretely, we define $\ell^{(m)}_1$ and $\ell^{(m)}_2$ as the values of $\ell^{(m)}_{\textup{max}}$ at $\bth^{(m)}=\bth^{(m)}_0$ (before solving $\cT_m$) and at $\bth^{(m)}=\bth^{(m)}_{\textup{opt}}$ (after solving $\cT_m$), respectively.
As $\ell^{(m)}_{\textup{max}}$ depends solely on the auxiliary task $\cT_m$, we assume that solving each $\cT_m$  reduces $\ell^{(m)}_{\textup{max}}$ by a consistent amount, $\ell^{(m)}_1 - \ell^{(m)}_2 = \Delta \ell$, for all $m$.
Consequently, the reduction in the upper bound in Eq.~(17) after solving $\cT_m$ is given by $(\Delta \ell) c_{M,m}$. 
Thus, a large (small) $c_{M,m}$ indicates that solving $\cT_m$ has a greater (lesser) reduction in the upper bound, resulting in a more (less) significant contribution to minimizing the empirical risk $R_{T_M}(h)$.

\subsubsection{Unitary learning task and Q-CurL game}
We consider the unitary learning task to verify the curriculum criteria based on $c_{M,m}$. We aim to optimize the parameters $\bth$ of a $Q$-qubit circuit $U(\bth)$, such that, for the optimized parameters $\bth_{\textrm{opt}}$, $U(\bth_{\textrm{opt}})$ can approximate an unknown $Q$-qubit unitary $V$ ($U, V \in \mathcal{U}(\mathbb{C}^{2^Q})$).

Our goal is to minimize the Hilbert-Schmidt (HS) distance between $U(\bth)$ and $V$, defined as
$
    C_{\textrm{HST}}(\bth) := 1 - \dfrac{1}{d^2}|\tr[V^{\dagger}U(\bth)]|^2,
$
where $d=2^Q$ is the dimension of the Hilbert space.
This HS distance is equivalent to the average fidelity between two evolved states under $U(\bth)$ and $V$ from the same initial state $\ket{\psi}$ drawn from the Haar uniform distribution of states:
\begin{align}
    C_{\textrm{HST}}(\bth) = \dfrac{d+1}{d}\bbE_{\ket{\psi}\sim \textrm{Haar}_n } \left[ 1 - |\braket{\psi| V^{\dagger}U(\bth)|\psi}|^2\right].
\end{align}
This suggests a QML-based approach to learn the target unitary $V$, where we can access a training data set consisting of input-output pairs of pure $Q$-qubit states $\cD_{\cQ}(N) = \{(\ket{\psi}_j, V\ket{\psi}_j)\}_{j=1}^N$ drawn from the distribution $\cQ$.
If we take $\cQ$ as the Haar distribution, we can instead train using the empirical loss:
\begin{align}\label{eqn:empirical:loss}
    C_{\cD_{\cQ}(N)}(\bth) := 1 - \dfrac{1}{N}\sum_{j=1}^N | \braket{\psi_j|V^{\dagger}U(\bth) |\psi_j}|^2.
\end{align}

\begin{figure}
		\includegraphics[width=8.7cm]{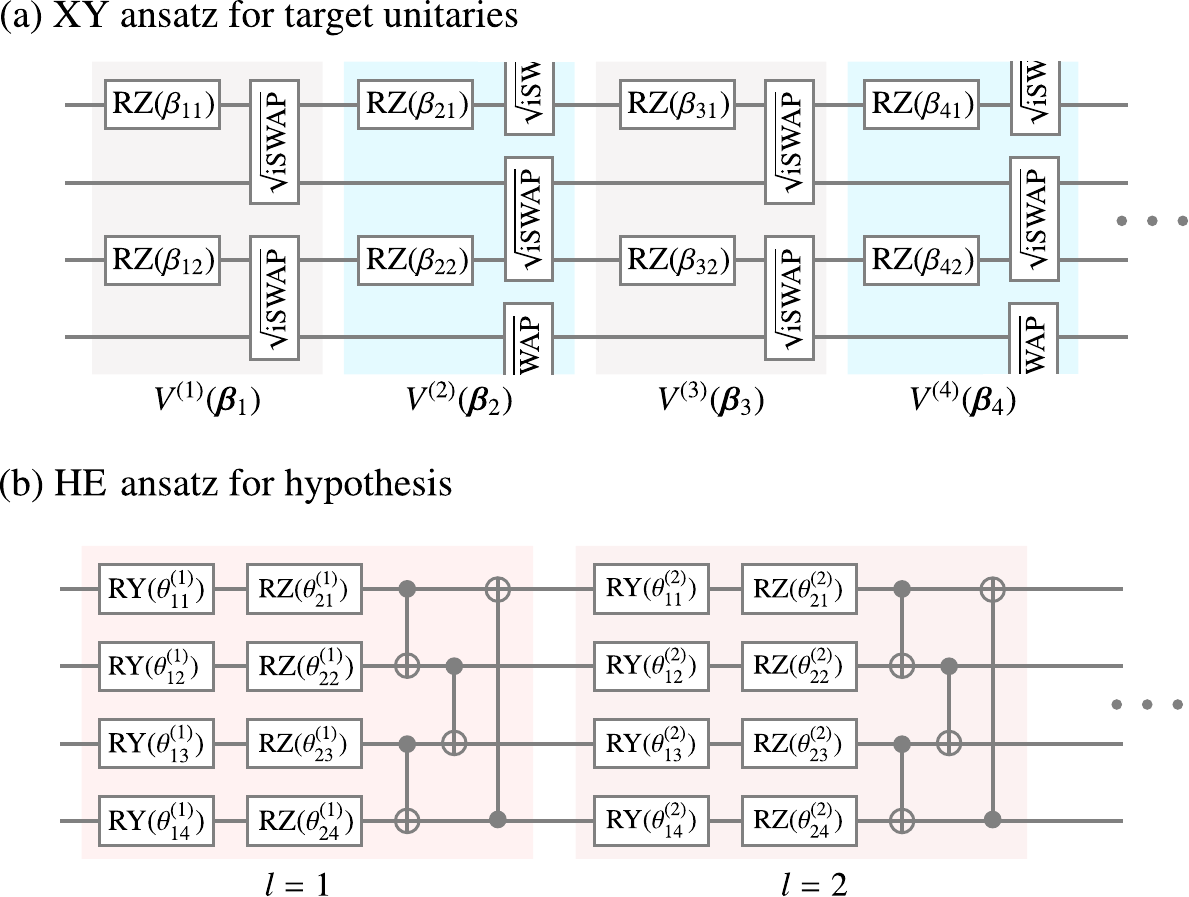}
		\protect\caption{(a) The XY ansatz, employed for constructing target unitaries in both the main and auxiliary tasks, is inspired by the XY model with periodic boundary conditions. It comprises single-qubit $RZ$ rotations applied to qubit indices $1, 3, \ldots$, and nearest-neighbor $\sqrt{\text{iSWAP}}$ gates, arranged with periodic boundary conditions. The placement of $\sqrt{\text{iSWAP}}$ gates in each layer $l$ varies depending on whether $l$ is odd or even. (b) The hardware-efficient (HE) ansatz, used to represent the hypothesis in the unitary learning task, consists of single-qubit RY and RZ rotations applied to all qubit indices, combined with nearest-neighbor CNOT gates, arranged with periodic boundary conditions.
		\label{fig:ansatz}}
\end{figure}

The parameterized ansatz $U(\bth)$ can be modeled as $U(\bth) = \prod_{l=1}^L U^{(l)}(\bth_l)$, consisting of $L$ repeating layers of unitaries. Each layer $U^{(l)}(\bth_l) = \prod_{k=1}^K \exp{(-i\theta_{lk} H_k)}$ is composed of $K$ unitaries, where $H_k$ are Hermitian operators, $\bth_l$ is a $K$-dimensional vector, and $\bth = \{\bth_1, \ldots, \bth_L\}$ is the $LK$-dimensional parameter vector.

We present a benchmark of Q-CurL for learning the approximation of the unitary dynamics of the spin-1/2 XY model with the following Hamiltonian (with the periodic boundary condition):
\begin{align}
H_{XY}=\sum_{j=1}^Q \left( \sigma_j^x\sigma_{j+1}^x + \sigma_j^y\sigma_{j+1}^y +  h_j\sigma_j^z \right),
\end{align}
where $h_j \in \bbR$ and $\sigma_j^x, \sigma_j^y, \sigma_j^z$ are the Pauli operators acting on qubit $j$.
This model is important in the study of quantum many-body physics, as it provides insights into quantum phase transitions and the behavior of correlated quantum systems.

To create the main task $\cT_M$ and auxiliary tasks, we represent the time evolution of $H_{XY}$ via the XY ansatz $V_{XY}$, which is similar to the Trotterized version of $\exp(-i\tau H_{XY})$~\cite{haug:2023:generalization}.
The target unitary for the main task consisting of $L_M=20$ repeating layers is defined as
\begin{align}
    V^{(M)}_{XY} = \prod_{l=1}^{L_M} V^{(l)}(\boldsymbol{\beta}_l) \prod_{l=1}^{L_F} V^{(l)}_{\textrm{fixed}},
\end{align}
where each layer $V^{(l)}(\boldsymbol{\beta}_l)$ includes parameterized z-rotations RZ (with assigned parameter $\boldsymbol{\beta}_l$) and non-parameterized nearest-neighbor $\sqrt{i\textup{SWAP}}=\exp(\frac{i \pi}{8} (\sigma_j^x \sigma_{j+1}^x + \sigma_j^y \sigma_{j+1}^y))$ gates [Fig.~\ref{fig:ansatz}(a)].
Here, $\boldsymbol{\beta}_l$ are initialized randomly from a uniform distribution over $[0, 1]$.
Additionally, we include the fixed-depth unitary $\prod_{l=1}^{L_F} V^{(l)}_{\textrm{fixed}}$ with $L_F=20$ layers at the end of the circuit $\prod_{l=1}^{L_M} V^{(l)}(\boldsymbol{\beta}_l)$ to increase expressivity.
Each $V^{(l)}_{\textrm{fixed}}$ shares the same ansatz structure as $V^{(l)}(\boldsymbol{\beta}_l)$, but with fixed $RZ$ gate angles assigned from a uniform distribution over $[0, 2\pi]$. The target unitary for the main task $\mathcal{T}_M$ includes $(L_M + L_F) \times \lceil Q/2 \rceil = 80$ RZ gate angles, where $L_M = 20$, $L_F = 20$, and $Q = 4$.

Similarity, we create the target unitary for the auxiliary tasks $\cT_m$ as 
\begin{align}
    V^{(m)}_{XY} = \prod_{l=1}^{L_m} V^{(l)}(\boldsymbol{\beta}_l)\prod_{l=1}^{L_F} V^{(l)}_{\textrm{fixed}},
\end{align}
with $L_m=1,2,\ldots, L_M-1$.
Therefore, the auxiliary task $\mathcal{T}_m$ consists of $L_m$ quantum circuits, $V^{(1)}(\boldsymbol{\beta}_1), \ldots, V^{(L_m)}(\boldsymbol{\beta}_{L_m})$, which form a subset of the $L_M$ quantum circuits $V^{(1)}(\boldsymbol{\beta}_1), \ldots, V^{(L_M)}(\boldsymbol{\beta}_{L_M})$ in the main task $\cT_M$. The parameter range for $\boldsymbol{\beta}_l$ is set to $[0, 1.0]$ instead of $[0, 2\pi]$ to ensure that the auxiliary tasks maintain a periodic similarity to the main task.

\begin{figure*}
		\includegraphics[width=18cm]{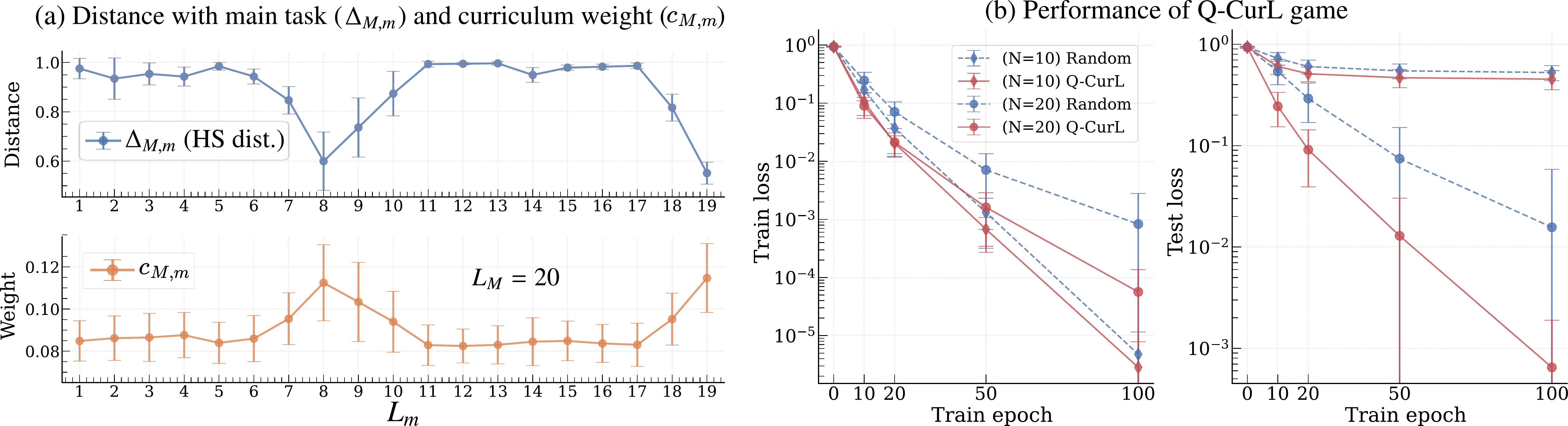}
		\protect\caption{(a) The curriculum weight (lower panel) and the Hilbert-Schmidt distance (upper panel) between the target unitary of the main task $\cT_M$ and the target unitary of the auxiliary task $\cT_m$. (b) The training loss and test loss for different training epochs and different numbers $N$ of training data in the Q-CurL game, considering both random and Q-CurL orders.
  The average and standard deviations are calculated over 100 trials.
		\label{fig:curr:game}}
\end{figure*}

\begin{figure*}
		\includegraphics[width=18cm]{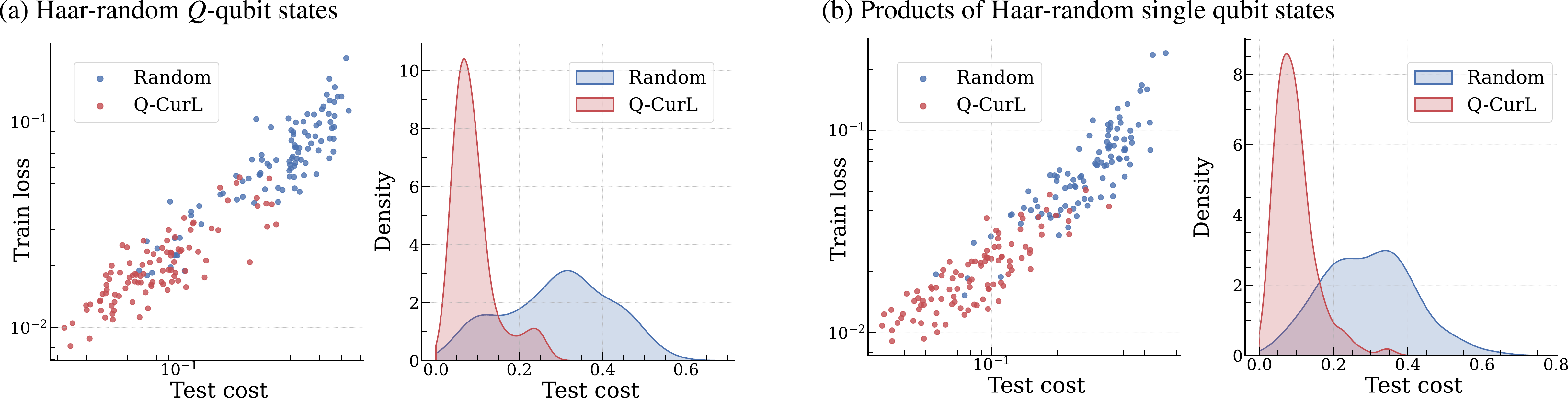}
		\protect\caption{The distribution and density of the training cost and test cost of the main task in the Q-CurL game, considering both random order and Q-CurL order based on the curriculum weights. Here, $N=20$ random input data are trained for 20 epochs with 100 trials of initial parameters in the model, and $N=20$ data are tested for each trained model.
  We consider two types of random input as (a) Haar-random $Q$-qubit states and (b) products of $Q$ Haar-random single-qubit states.
		\label{fig:curr:dist}}
\end{figure*}

In our experiments, we consider the unitary learning with $Q=4$ qubits via the hardware efficient (HE) ansatz $U_{\textrm{HEA}}(\bth)$. This ansatz comprises multiple blocks, where each block consists of single-qubit operations spanned by SU(2) on all qubits and two-qubit controlled-X entangling gates~\cite{barkoutsos:2018:twolocal} repeated for all pairs of neighbor qubits. Here, we use rotation operators of Pauli Y and Z as single qubit gates. Mathematically, $U_{\textrm{HEA}}(\bth)$ is defined as follows:
\begin{align}~\label{HEA:circuit}
    U_{\textrm{HEA}}(\bth) = \prod_{l=1}^{L_E} \left( \prod_{q=1}^Q \left[ U_{R}^{q,l}(\bth^{(l)}) \right] \times U_{\text{Ent}} \right),
\end{align}
with $Q$ qubits consisting of $L_E$ entangling gates $U_{\text{Ent}}$ alternating with rotation gates on each qubit.
Here, we use $U_R(\bth) = R_Y(\bth_1)R_Z(\bth_2)$, and  $U_{\text{Ent}}$ is composed of CNOT gates placed in linear with indexes $(q,q+1)$ of qubits, arranged with periodic boundary conditions [Fig.~\ref{fig:ansatz}(b)]. The number of parameters in this circuit is $2QL_E$.
We configure the HE ansatz with $L_E = 40$ layers to achieve high expressivity, resulting in 320 parameters for $Q = 4$ qubits. These parameters are initialized randomly from a uniform distribution over $[0, 2\pi]$.

Figure~\ref{fig:curr:game}(a) depicts the average HS distance over 100 trials of $\bbta_l$ and $V^{(l)}_{\textrm{fixed}}$ between the target unitary of each auxiliary task $\cT_m$ (with $L_m$ layers) and the main task $\cT_M$.
The Hilbert-Schmidt (HS) distance $\Delta_{M,m}$ represents the distance between the unitaries $V^{(m)}_{XY}$ and $V^{(M)}_{XY}$. 
It is calculated as follows, assuming the explicit forms of $V^{(m)}_{XY}$ and $V^{(M)}_{XY}$ are known: 
\begin{align}
    \Delta_{M,m}:= 1 - \dfrac{1}{d^2}\left|\tr[V^{(m)\dagger}_{XY}V^{(M)}_{XY}]\right|^2,
\end{align}
where $d=2^Q$ denotes the dimension of the Hilbert space.
The HS distance here depends solely on the design of the target unitaries for the auxiliary tasks and the main task, and is independent of the trainable ansatz $U_{\textup{HEA}}(\boldsymbol{\theta})$. Thus, it offers insight into which auxiliary task is most similar to the main task by comparing the distances $\Delta_{M,1}, \Delta_{M,2}, \ldots, \Delta_{M,M-1}$.
As shown in Fig.~\ref{fig:curr:game}(a), the HS distances at auxiliary tasks $\mathcal{T}_8$ and $\mathcal{T}_{19}$ are approximately 0.6, which are considerably smaller than those at other auxiliary tasks (approximately 1.0).
As the HS distance cannot be computed in advance, we rely on the curriculum weight to guide the design of the curriculum.

We plot the curriculum weight $c_{M,m}$ in Fig.~\ref{fig:curr:game}(a) calculated in Eq.~\eqref{eqn:curr:weight} with the basis functions form in Eq.~\eqref{eqn:basis:trace}.
Here, we use $N=20$ Haar random states for input data $\bx_i^{(m)}$ in each task $\cT_m$.
As depicted in Fig.~\ref{fig:curr:game}(a), $c_{M,m}$ can capture the similarity between two tasks, as higher weights imply smaller HS distances.

Next, we propose a Q-CurL game to further examine the effect of  Q-CurL. In this game, Alice has an ML model $\cM(\bth)$ to solve the main task $\cT_M$, but she needs to solve all the auxiliary tasks $\cT_1, \ldots, \cT_{M-1}$ first. We assume the data forgetting in task transfer, meaning that after solving task $A$, only the trained parameters derived from task $A$ are transferred as the initial parameters for task $B$. 
The Q-CurL framework provides an algorithm to determine an efficient order of auxiliary tasks to facilitate solving the main task.
We propose the following greedy algorithm to decide the curriculum order $\cT_{i_1}\to \cT_{i_2} \to \ldots \to \cT_{i_M=M}$ before training.
Starting $\cT_{i_M}$, we find the auxiliary task $\cT_{i_{M-1}}$ ($i_{M-1} \in \{1,2,\ldots,M-1\}$) with the highest curriculum weights $c_{i_M,i_{M-1}}$.
Similarity, to solve $\cT_{i_{M-1}}$, we find the corresponding auxiliary task $\cT_{i_{M-2}}$ in the remaining tasks with the highest $c_{i_{M-1},i_{M-2}}$, and so on.
Here, curriculum weights $c_{i_{k}, i_{k-1}}$ are calculated similarly to Eq.~\eqref{eqn:curr:weight}.

Figure~\ref{fig:curr:game}(b) depicts the training and test loss of the main task $\cT_M$ (see Eq.~\eqref{eqn:empirical:loss}) for different training epochs and numbers of training data over 100 trials of parameters' initialization.
In each trial, $N$ Haar random states are used for training, and 20 Haar random states are used for testing.
With a sufficient amount of training data ($N=20$), introducing Q-CurL can significantly improve the trainability (lower training loss) and generalization (lower test loss) when compared with random order in Q-CurL game.
Even with a limited amount of training data ($N=10$), when overfitting occurs, Q-CurL still performs better than the random order.

Figure~\ref{fig:curr:dist} depicts the distribution and density of the train loss and test loss of the main task in the Q-CurL game, comparing the Q-CurL order with a random order. Here, $N=20$ random input data are trained for 20 epochs with 100 trials of initial parameters in the model, and $N=20$ data are tested for each trained model.
We consider two types of random inputs as (a) Haar-random $Q$-qubit states [Fig.~\ref{fig:curr:dist}(a)] and (b) products of $Q$ Haar-random single-qubit states [Fig.~\ref{fig:curr:dist}(b)].
In both types of input states, the order in solving the Q-CurL game derived via the task-based Q-CurL method outperforms the performance when considering the random order.

The Q-CurL game setting and the heuristic greedy algorithm discussed here demonstrate the usefulness of using curriculum weight to decide the curriculum order. We can further explore several variations of the Q-CurL game.
For instance, instead of using the test loss $\cL^{(M)}_t$ of the main task $\cT_M$ as the evaluation metric for the curriculum order $\cT_{i_1}\to \cT_{i_2} \to \ldots \to \cT_{i_M=M}$, one could consider minimizing the total test loss $\sum_{k=2}^M \cL^{(i_k)}_t$ . This approach would lead to a heuristic algorithm aimed at maximizing the total curriculum weights $\sum_{k=2}^M c_{i_k,i_{k-1}}$.
Another variation is to consider the task difficulty perspective. For example, we could set the first task to be solved initially (as we know it is easy to solve, or we already have a trained model) and then determine an optimal task order that smoothly transitions from the first task to the main task.

\subsection{Data-based Q-CurL}
We present a form of data-based Q-CurL that dynamically predicts the easiness of each sample at each training epoch, such that easy samples are emphasized with large weights during the early stages of training and conversely. 
Remarkably, it does not involve pre-training or additional training data, thereby avoiding any increase in quantum resource requirements.

Apart from improving generalization, data-based Q-CurL offers resistance to noise.
This feature is particularly valuable in QML, where clean annotated data are often costly while noisy data are abundant.
Existing QML models can accurately fit corrupted labels in the training data but often fail on test data~\cite{fuster:2024:natcom}. We demonstrate int the quantum phase recognition task that data-based Q-CurL enhances robustness by dynamically weighting the difficulty of fitting corrupted labels.


\subsubsection{Dynamical learning schedule}\label{sec:dyn:loss}
In the procedure without using the Q-CurL, we use the conventional loss as the empirical risk $\hat{R}(h) = \dfrac{1}{N}\sum_{i=1}^N \ell_i$ for the training and testing phase.
In data-based Q-CurL, inspired by the confidence-aware techniques in classical ML~\cite{novot:2018:CVPR,saxena:2019:dataparm:NIPS,superloss:2020:NIPS}, 
we modify the conventional loss to the following form of the dynamical loss function:
\begin{align}\label{eqn:mod:loss}
        \hat{R}(h, \bw) = \dfrac{1}{N}\sum_{i=1}^N \left( ( \ell_i - \eta )e^{w_i} + \gamma w^2_i \right). 
\end{align}
Here, $\bw=(w_1,\ldots,w_N)$ and $w^2_i$ is the regularization term controlled by the hyper-parameter.
The threshold $\eta$ distinguishes easy and hard samples with $e^{w_i}$ emphasizing the loss $\ell_i \ll \eta$ (easy sample) or the loss $\ell_i \gg \eta$ (hard samples, such as data with corrupted labels).
The optimization is reduced to
\begin{align}
\textup{min}_{\bth}\textup{min}_{\bw}\hat{R}(h, \bw),
\end{align}
where $\bth$ is the parameter of the hypothesis $h$. Here, $\textup{min}_{\bw}\hat{R}(h, \bw)$ is decomposed at each loss $\ell_i$ and solved without quantum resources as 
\begin{align}\label{eqn:loss:weight}
w_i = \textup{argmin}_w(l_i-\eta)e^{w} + \gamma w^2.    
\end{align}
To control the difficulty of the samples, in each training epoch, we set $\eta$ as the average value of all $\ell_i$ obtained from the previous epoch. Therefore, $\eta$ adjusts dynamically in the early training stages but stabilizes near convergence.

In Appendix~\ref{sec:loss:func}, we present the details of solving Eq.~\eqref{eqn:loss:weight}.
Given the solution of Eq.~\eqref{eqn:loss:weight}, by controlling the sign of $\gamma$, the dynamical loss can be used to prioritize emphasizing the small losses (with $\gamma > 0$) or the large losses (with $\gamma < 0$). We define these two scenarios as \textit{easy Q-CurL} and \textit{hard Q-CurL}, respectively.

The easy Q-CurL aligns with the classical curriculum learning context~\cite{superloss:2020:NIPS}. This approach is particularly beneficial when hard samples, such as those with noisy labels, could mislead the optimization process. However, the hard Q-CurL can be advantageous in scenarios where hard samples such as the complex quantum data are crucial for guiding the model to extract essential features without being distracted by irrelevant ones. We will provide an example of this interesting scenario at the end of the next subsection.


\begin{figure}
		\includegraphics[width=8.7cm]{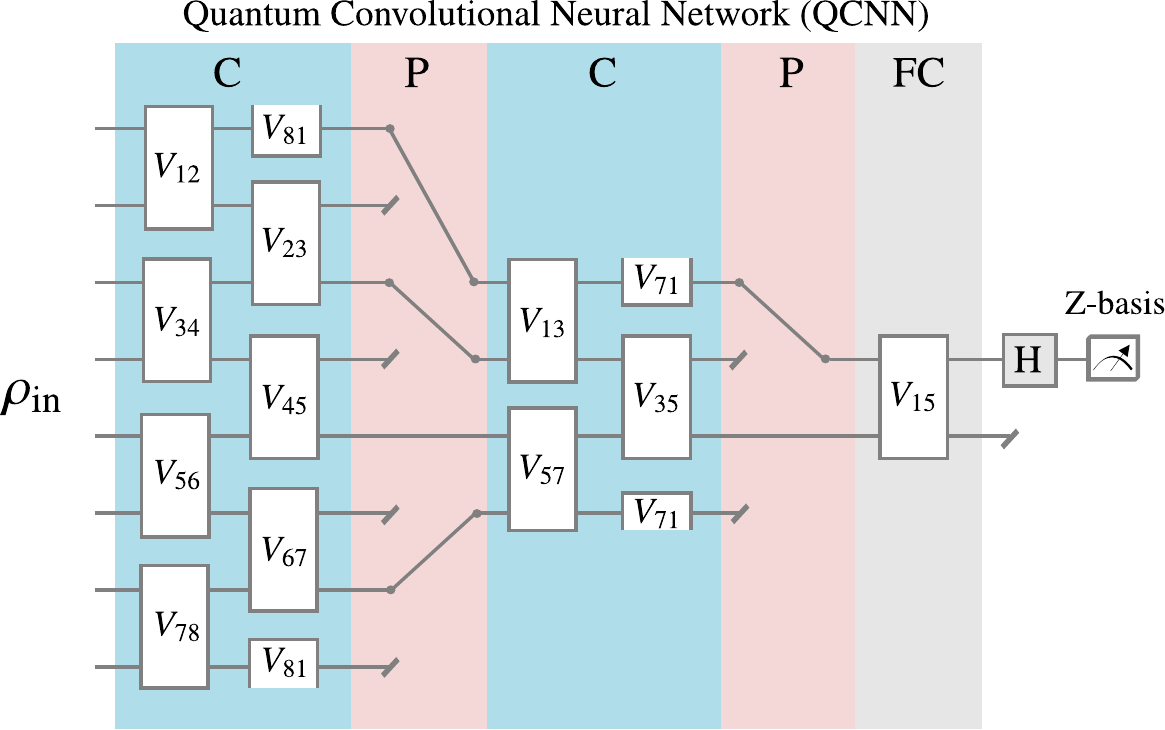}
		\protect\caption{The schematic diagram for the QCNN~\cite{cong:2019:QCNN} used in our quantum phase recognition task. Here, C, P, and FC represent convolutional, pooling, and fully connected layers, respectively. In the convolutional layer, local unitaries $V_{kl}$ are applied to pairs of neighboring qubits $(k,l)$ in the illustrated order, excluding those previously discarded, under periodic boundary conditions. 
        All $V_{kl}$ in the same layer share the same parameters.
        In the pooling layer, qubits with even indices among the remaining qubits are discarded. This sequence of alternating convolutional and pooling layers ends with a fully connected layer, which operates as a single convolutional operator on the remaining qubits. Finally, we apply the Hadamard gate to the last remaining qubit and then perform a measurement in the Z-basis to classify the input quantum data $\rho_{\textup{in}}$.
		\label{fig:qcnn:ansatz}}
\end{figure}

\begin{figure*}
		\includegraphics[width=18cm]{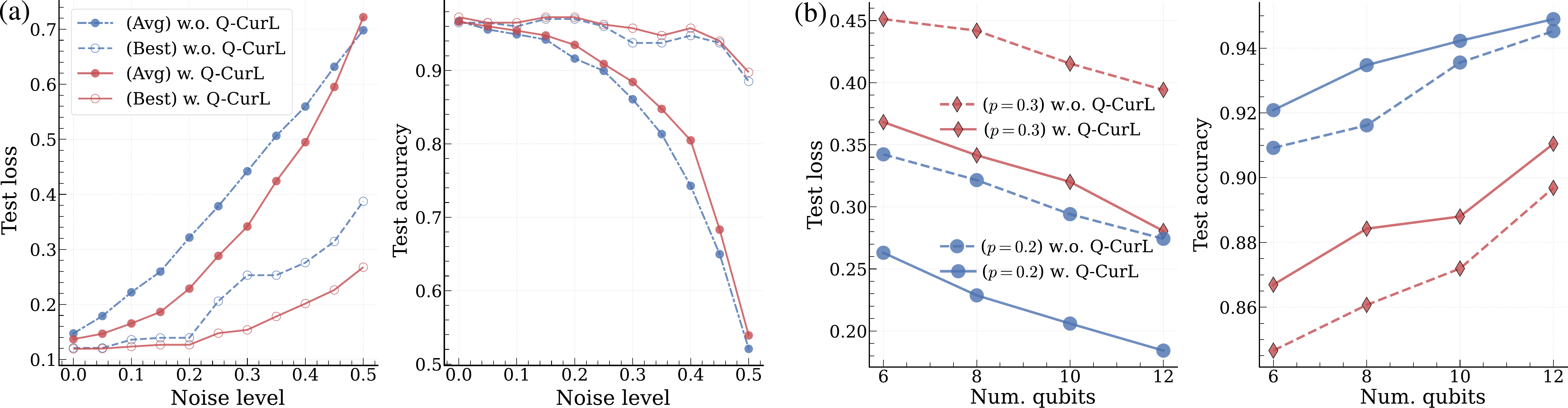}
		\protect\caption{(a) The test loss and accuracy of the trained QCNN (with and without using the data-based Q-CurL) in the quantum phase recognition task with 8 qubits under varying noise levels in corrupted labels. Here, the average and the best performance over 50 trials are plotted. 
        (b) The test loss (left panel) and test accuracy (right panel) of the trained QCNN on the quantum phase recognition task with (solid lines) or without (dotted lines) using the data-based Q-CurL over different numbers of qubits. Here, we consider two different noise levels in the corrupted training labels: $p=0.2$ (blue) and $p=0.3$ (red).
		\label{fig:qcurr:costgen}}
\end{figure*}

\subsubsection{Quantum phase recognition task}
We apply the data-based Q-CurL to the quantum phase recognition task investigated in Ref.~\cite{cong:2019:QCNN} to demonstrate that it can improve the generalization of the learning model. Here, we consider a one-dimensional cluster Ising model with open boundary conditions, whose Hamiltonian with $Q$ qubits is given by
\begin{align}~\label{eqn:Hamiltonian:open}
    H = - \sum_{i=1}^{Q-2}\sigma^z_i \sigma^x_{i+1}\sigma^z_{i+2} - h_1\sum_{i=1}^Q \sigma^x_i - h_2\sum_{i=1}^{Q-1}\sigma^x_{i}\sigma^x_{i+1}.
\end{align}
Depending on the coupling constants $(h_1, h_2)$, the ground state wave function of this Hamiltonian can exhibit multiple states of matter, such as the symmetry-protected topological phase (SPT phase), the paramagnetic state, and the anti-ferromagnetic state.

We employ the quantum convolutional neural network (QCNN) model~\cite{cong:2019:QCNN} to determine the matter phase of quantum states. Inspired by classical convolutional neural networks, the QCNN model consists of convolutional, pooling, and fully connected layers. The convolutional layers use local unitary gates to extract local features from the input data, while the pooling layers reduce the number of qubits. This alternation of layers ends in a fully connected layer that functions as a single convolution operator on the remaining qubits, providing an output through the measurement of the final qubit.
The QCNN is governed by variational parameters that are optimized to classify training data accurately.
In our implementation, the convolutional and fully connected layers are constructed using the Pauli decomposition of two-qubit unitary gates $V_{kl}$, expressed as $V_{kl}=\prod_{j=1}^{15}e^{-i\theta_jP_j}$ with 15 parameters, where $\{P_j\}$ are the Pauli operators for two qubits, excluding the identity matrix. 
Here, $V_{kl}$ applies to pairs of neighboring qubits $(k,l)$ with the order illustrated in Fig.~\ref{fig:qcnn:ansatz}, excluding those previously discarded.
Each layer utilizes the same parameters for all unitary gates. In the pooling layer, qubits with even indices among the remaining qubits are discarded. Before measuring the output, we apply the Hadamard gate to the remaining qubit and then perform a measurement in the Z-basis.
In this QCNN scheme, the number of independent parameters is given by $15\times \lceil \log_2(Q) \rceil$, where $\lceil \log_2(Q) \rceil$ represents the number of convolutional and fully connected layers.

For each training quantum data $\ket{\psi_i}$ and its corresponding label $y_i$, the QCNN produces the output $q_i$ ($-1\leq q_i \leq 1$). The single loss $\ell_i$ is defined using the binary cross-entropy (BCE) loss as follows:
\begin{align}
    \ell_i = -y_i\log(\hat{y}_i) - (1.0-y_i)\log(1.0-\hat{y}_i),
\end{align}
where $\hat{y}_i=\textup{sigmoid}(\mu q_i)$. 
Here, we consider the scaling output with the coefficient $\mu=1.0$.
The label is predicted as $0$ if $\hat{y}_i < 0.5$ and $1$ if $\hat{y}_i \geq 0.5$.
In the procedure without using the Q-CurL, we use the conventional loss $\hat{R}(h) = \dfrac{1}{N}\sum_{i=1}^N \ell_i$ for the training.
We also use $\hat{R}(h)$ to evaluate the generalization on the test data set.

Similar to the setup in Ref.~\cite{cong:2019:QCNN}, we generate a training set of 40 ground state wave functions corresponding to $h_2=0$ and $h_1$ sampled at equal intervals in [0.0, 1.6]. The state is analytically solvable for these parameter choices, and this solution is used to label the training dataset (0 for the paramagnetic or antiferromagnetic phase and 1 for the SPT phase). The ground truth phase boundaries, which separate the two phases, are determined using DMRG simulations. Based on these boundaries, we also create a test dataset of 400 ground state wave functions corresponding to $h_2 \in$ \{0.8439, 0.6636, 0.5033, 0.3631, 0.2229, 0.09766, -0.02755, -0.1377, -0.2479, -0.3531\}, and $h_1$ sampled 40 times at equal intervals in [0.0, 1.6]. The optimization is performed by the Adam method with a learning rate of 0.001 and 500 epochs of training.

In our experiment, we consider the scenario of fitting corrupted labels. Given a probability $p$ ($0\leq p \leq 1$) representing the noise level, the true label $y_i \in\{0, 1\}$ of quantum state $\ket{\psi_i}$ is transformed to the corrupted label $1-y_i$ with probability $p$, while it remains the true label with probability $1-p$.

Figure~\ref{fig:qcurr:costgen}(a) illustrates the performance of a trained QCNN on test data across various noise levels. There is a minimal difference at low noise levels, but as noise increases, conventional training fails to generalize effectively. Introducing data-based Q-CurL in training (red lines) reduces test loss and improves test accuracy compared to the conventional method (blue lines).

\begin{figure*}
		\includegraphics[width=17cm]{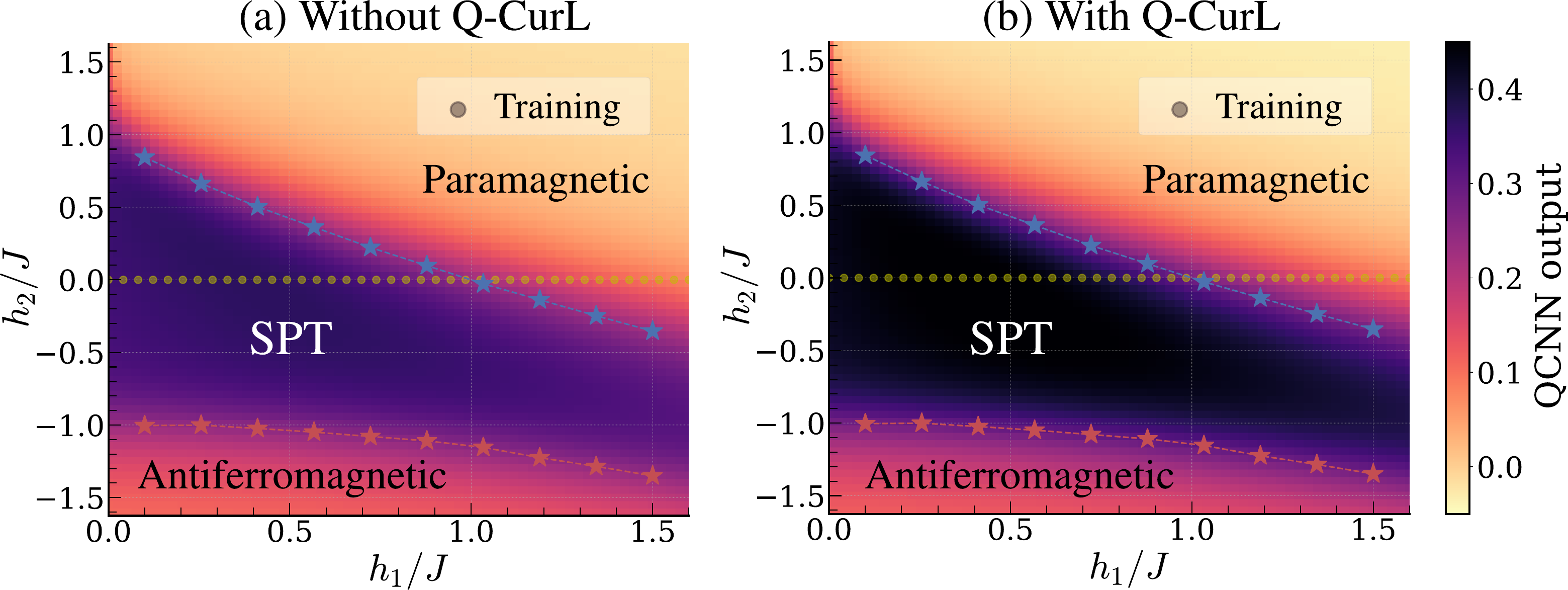}
		\protect\caption{\small The heatmap showing the average output of the QCNN over 50 trials of initial parameters in the cases of (a) without Q-CurL and (b) with Q-CurL for combinations of $(h_1/J, h_2/J)$ and the probability of corrupted label is $p=0.3$ in the quantum phase recognition task with $Q=8$ qubits. The dotted points indicate the data points used during training, while the blue and red lines with star markers highlight the true boundaries between the SPT phase, the paramagnetic phase, and the antiferromagnetic phase. Introducing Q-CurL enhances the separation between the SPT phase and others, with higher values for the SPT phase and lower values for other phases.
		\label{fig:phase:vis}}
\end{figure*}

Figure~\ref{fig:qcurr:costgen}(b) illustrates the average performance of trained QCNN on test data with noise levels $p=0.2, 0.3$ in corrupted training labels over different numbers of qubits. Introducing data-based Q-CurL (solid lines) in the training process reduces the test loss and enhances testing accuracy compared to the conventional training method (dotted lines). 
We note that introducing noise in the training labels leads to worse generalization in the system with fewer qubits.
The small QCNN model struggles to extract the correct phase of the quantum data with limited information. However, as the number of qubits increases, more information is provided in the quantum wave functions for the QCNN to extract, thereby improving the robustness in phase detection tasks.

In Fig.~\ref{fig:phase:vis}, we present a heatmap showing the average QCNN output over 50 trials with different initial parameters, comparing cases (a) without Q-CurL and (b) with (easy) Q-CurL, across combinations of $(h_1/J, h_2/J)$ with a corrupted label probability of $p=0.3$ and $Q=8$ qubits. In this experiment, we consider the same Hamiltonian form in Eq.~\eqref{eqn:Hamiltonian:open} but with periodic boundary conditions. Additionally, we employ in QCNN the following ansatz circuit for the convolutional and fully connected layers with the depth $d_c=5$:
\begin{align}
    V = \prod_{i=1}^{d_c} U_{1i}(\bth^{(1i)}) U_{2i}(\bth^{(2i)}).
\end{align}
Here, $U_{1i}(\bth^{(1i)})$ is the product of rotation gates $\prod_{j=1}^3e^{-i\theta^{(1i)}_{j} P^{(1)}_j}$ applied to each single qubit $k$ ($1\leq k\leq Q$), where $\{P_j^{(1)}\}$ are the Pauli operators for single qubit, excluding the identity operator. Here, all $k$ in the same layer share the same parameters in $U_{1i}(\bth^{(1i)})$. Similarly, $U_{2i}(\bth^{(2i)})$ is the product of two-neighbor qubit gates  $\prod_{j=1}^{15}e^{-i\theta^{(2i)}_{j} P^{(2)}_j}$ on two qubits $(k,k+1)$ with the periodic boundary condition, where $\{P_j^{(2)}\}$ are the Pauli operators for two qubits, excluding the identity operator. Here, all pairs $(k, k+1)$ in the same layer share the same parameters in $U_{2i}(\bth^{(2i)})$. 
Therefore, this QCNN has $d_c \times (3+15) \times \lceil \log_2(Q) \rceil$ independent parameters, where $\lceil \log_2(Q) \rceil$ represents the number of convolutional and fully connected layers. For $Q=8$ and $d_c=5$, the number of independent parameters is $270$.
 
We also use the following form of the single loss $\ell_i$:
\begin{align}
    \ell_i = -s(y_i)\log(\hat{y}_i) - (1.0-s(y_i))\log(1.0-\hat{y}_i),
\end{align}
where $\hat{y}_i=\textup{sigmoid}(5.0 q_i)$ is the post-processing of the QCNN's output for faster convergence of the loss function. Here, $s(y_i)$ transforms the label $y_i$ to control for the range of QCNN's output during training. In previous experiments, we set $s(y_i)$ as an identity map $s(y_i)=y_i$. However, with random initialization, the QCNN output $q_i$ remains close to zero, making post-processed value $\hat{y}_i$ approximately 0.5. To accelerate optimization, we modify the transformation such that $\hat{y}_i$ approaches 1.0 for data in the SPT phase, while data in other phases remain near 0.5. Specifically, we set $s(y_i)=0.5$ for $y_i=0$ and $s(y_i)=1.0$ for $y_i=1$.

We explain the usage of data in training and evaluating. The dotted points in Fig.~\ref{fig:phase:vis} indicate the data points used during training. The blue and red dotted lines with star markers highlight the true boundaries between the SPT phase (middle), the paramagnetic phase (upper), and the antiferromagnetic phase (lower).
For the test dataset, we sampled $h_1$ and $h_2$ 64 times at equal intervals within the ranges [0.0, 1.6] and [-1.6, 1.6], respectively. 
Fig.~\ref{fig:phase:vis} depicts that introducing Q-CurL enhances the separation between the SPT phase and other quantum phases, with lower values for the paramagnetic phase and antiferromagnetic phase, and higher values for the SPT phase.
Therefore, Q-CurL offers more reliable insights into the use of QML for understanding physical systems.

\subsubsection*{Curriculum learning with easy or hard samples?}

\begin{figure}
		\includegraphics[width=8.7cm]{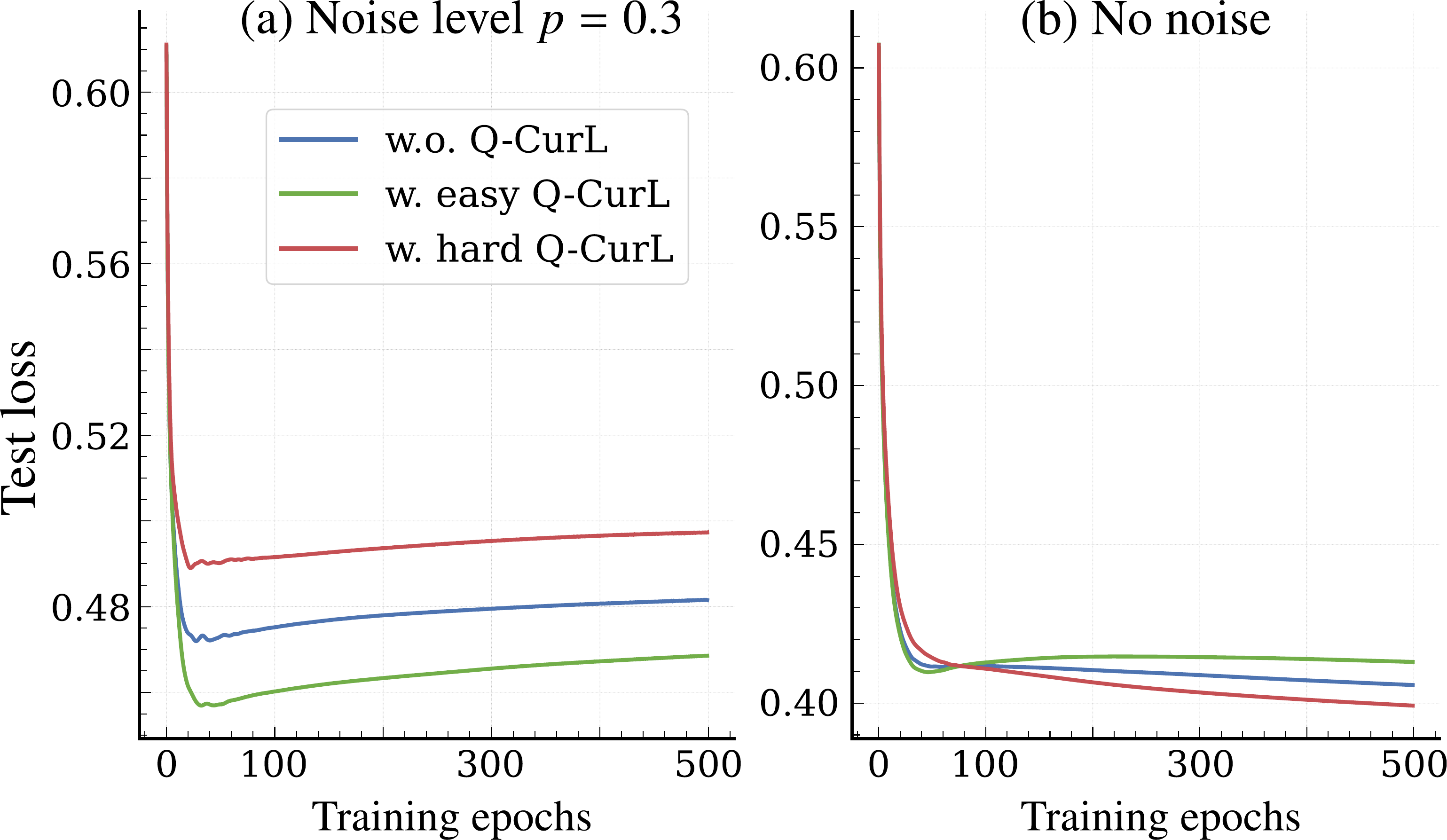}
		\protect\caption{The test loss in quantum phase detection task with $n=8$ qubits for different training loss types: without Q-CurL (blue), with easy Q-CurL (green, $\gamma=1.0$), and with hard Q-CurL (red, $\gamma=-1.0$). The losses are averaged over fifty experimental runs with different initializations of QCNN parameters. Two scenarios are considered: (a) data containing corrupted labels with a probability (noise level) of $p=0.3$, and (b) data without corrupted labeling (no noise).
		\label{fig:hard:examples}}
\end{figure}

At the end of Section~\ref{sec:dyn:loss}, we mentioned the easy Q-CurL and hard Q-CurL losses and identified scenarios where these different types of losses can be effectively utilized.
In revising our manuscript, we came across Ref.~\cite{erik:2024:complexity}, which appeared on arXiv after our paper. This reference presents a numerical result indicating that, in the task of quantum phase recognition, prioritizing harder data points early in the training process can lead to superior performance compared to traditional training methods. While this is an intriguing result that needs further investigation into the underlying reasons, we present a comparison between the easy Q-CurL and hard Q-CurL losses across different situations.

We employ the same setup as the experiment that produced the results in Fig.~\ref{fig:phase:vis}. In Fig.~\ref{fig:hard:examples}, we plot the test loss for different training loss types: without Q-CurL (blue), with easy Q-CurL (green, $\gamma=1.0$), and with hard Q-CurL (red, $\gamma=-1.0$). The test loss curves are averaged over fifty experimental runs with different initializations of QCNN parameters. For data with corrupted labeling, when hard data includes incorrect labels, it should not contribute to optimization. This is confirmed in Fig.~\ref{fig:hard:examples}(a), where the hard Q-CurL results in the highest test loss, while the easy Q-CurL achieves the lowest test loss.
Conversely, for data without corrupted labeling [Fig.~\ref{fig:hard:examples}(b)], during the early stages of training, easy Q-CurL may reduce the test loss more quickly than hard Q-CurL. However, with sufficient training epochs, hard Q-CurL achieves the best performance among these methods, without increasing the test loss as optimization continues.
Exploring why hard Q-CurL outperforms easy Q-CurL and traditional training methods without Q-CurL remains an interesting topic, particularly for the phase detection task.

\section{Conclusion and Discussion}
The proposed Q-CurL framework can enhance training convergence and generalization in QML with quantum data. 
A natural question arises: instead of applying the Q-CurL framework, could one employ classical curriculum learning techniques on classical representations of quantum data? Processing quantum data by first obtaining its classical representation and then applying classical ML shows promise for near-term applications. However, a significant bottleneck remains: efficiently constructing this classical representation without losing the intrinsic quantum characteristics of the data poses an ongoing challenge. 
Furthermore, this approach requires the development of a specialized interface for such classical representation data, beyond simply applying conventional ML methods.
In contrast, our approach, which operates directly with quantum data, circumvents this issue and may provide a practical advantage by preserving the quantum nature of the input.

It is beyond the scope of this study to compare the performance of quantum models directly against classical models.
Therefore, Q-CurL is not intended as a direct competitor to established classical methods. Rather, it serves as a framework to enhance existing quantum learning algorithms. The introduction of a curriculum learning weight within a task-based approach, combined with the exploration of emphasizing easy or hard samples through a dynamical loss function, provides the QML community with actionable strategies to improve performance. These contributions are particularly valuable for applications in chemistry and physics, where advancements in QML techniques can deliver substantial practical benefits.

Future research should investigate whether Q-CurL can be designed to improve trainability in QML, particularly by avoiding the barren plateau problem. For instance, curriculum design is not limited to tasks and data but can also involve the progressive design of the loss function. Even when the loss function of the target task, designed for infeasibility in classical simulation to achieve quantum advantage~\cite{cerezo:2023:BP:sim,fuster:2024:VQA:dequan}, is prone to the barren plateau problem, a well-designed sequence of classically simulable loss functions can be beneficial. Optimizing these functions in a well-structured curriculum before optimizing the main function may significantly improve the trainability and performance of the target task.

\begin{acknowledgments}
The authors acknowledge Koki Chinzei and Yuichi Kamata for their fruitful discussions. 
Special thanks are extended to Koki Chinzei for his valuable comments on the variations of the Q-CurL game.
\end{acknowledgments}

\newpage
\appendix

\section{Minimax framework for transfer learning in unitary learning task}

The task-based Q-CurL framework leaves several fundamental questions regarding the implementation of transfer learning algorithms from an auxiliary task to a main task. For example, what is the best accuracy that can be achieved through any transfer learning algorithm? How does this accuracy depend on the transferability between tasks? How does the accuracy of the main task in transfer learning scale with the amount of data in both the auxiliary and main tasks? In this section, we formulate the general minimax framework for transfer learning within the task-based Q-CurL framework. Specifically, for the unitary learning task, we map the minimax lower bounds for transfer learning with parameterized quantum circuits to the derivation of minimax lower bounds in transfer learning for linear regression problems. However, the detailed form of this bound is left for future research.


Here, we focus on the unitary learning task. We assume the presence of an auxiliary task $\cT_m$ and a main task $\cT_M$, with target unitaries $V_m$ and $V_M$ ($V_m, V_M \in \mathcal{U}(\mathbb{C}^{2^Q})$), respectively. 
In the auxiliary task, we can access a training data set $\cA_m$ consisting of $N_m$  input-output pairs of $Q$-qubits states as $\cA_m=\left\{\left(\ket{\psi^{(m)}_j}, \cE(V_m\ket{\psi^{(m)}_j}, \epsilon^{(m)}_j) \right)\right\}_{j=1}^{N_m}$, where $\cE$ is a quantum noise channel applied to the pure state $V_m\ket{\psi_j}$ with noise variable $\epsilon^{(m)}_j$. Here, $\epsilon^{(m)}_j=0$ implies that the identity operator $\cE$ is applied to the quantum state. We assume that the output of $\cE$ is represented in the form of a density matrix.
Similarly, in the main task $\cT_M$, we have access to a training dataset $\cA_M$ consisting of $N_M$ input-output pairs of $Q$-qubit states, denoted as $\cA_M=\left\{\left(\ket{\psi^{(M)}_j}, \cE(V_M\ket{\psi^{(M)}_j}, \epsilon^{(M)}_j) \right)\right\}_{j=1}^{N_M}$.
Furthermore, we assume that the input data $\ket{\psi^{(m)}_j}$ and $\ket{\psi^{(M)}_j}$ for both tasks are drawn from the same distribution $\cQ$, and each noise variable $\epsilon_j$ is drawn from a normal distribution $\cN(0, \sigma^2)$ with mean zero and variance $\sigma^2$.

With the notion of the HS distance between two unitaries as $\textup{HS}(U, V) = 1 - \dfrac{1}{d^2}|\tr[V^{\dagger}U(\bth)]|^2$ ($d=2^Q$), we formally define the transfer class of pairs of unitaries as
\begin{align}
    \cP_{\Delta} = \{(U, V) | U, V \in \mathcal{U}(\mathbb{C}^{d}) ; \textup{HS}(U, V) \leq \Delta  \}.
\end{align}

In a transfer learning problem, we are interested in using both auxiliary and main training data to find an estimate of the target unitary $V_M$ for the main task with a small generalization error. In the minimax approach, $V_M$ is chosen in an adversarial way, and the goal is to find and estimate $U_M$ that achieves the smallest worst-case target generalization risk (over the distribution $\cQ$):
\begin{align}\label{eqn:sup:risk}
    \textup{sup}_{\textup{transfer class}}\bbE_{\textup{auxiliary and main samples}}\left[ \bbE_{\cQ} \textup{loss} \right].
\end{align}

Formally, given an input data $\ket{\psi^{(M)}_j}\sim \cQ$, the loss induced by this data and the estimated $U_M(\bth_M)$ is expressed as
\begin{widetext}
\begin{align}\label{eqn:loss:transfer}
\ell_j(\bth_M) = 1.0 - \bra{\psi^{(M)}_j} U^{\dagger}_M(\bth_M) \cE(V_M\ket{\psi^{(M)}_j}, \epsilon^{(M)}_j)  U_M(\bth_M) \ket{\psi^{(M)}_j}.
\end{align}    
\end{widetext}
Then, minimizing Eq.~\eqref{eqn:sup:risk} can be written as the following transfer learning minimax risk:
\begin{align}\label{eqn:transfer:minimaxrisk}
    \cR_M(\cP_{\Delta}) := \textup{inf}_{\bth_M} \textup{sup}_{(V_m, V_M) \in \cP_{\Delta}}\bbE_{\cA_m}\bbE_{\cA_M}\left[ \bbE_{\cQ} \ell_j(\bth_M) \right].
\end{align} 

We would like to know a lower bound on the transfer learning minimax risk in Eq.~\eqref{eqn:transfer:minimaxrisk} to characterize the fundamental limits of transfer learning.
We note that this problem is very similar to the minimax framework in linear regression problems~\cite{kalan:2020:minimax}. 
Generally, any $Q$-qubit density matrix $\rho$ has a unique representation as 
\begin{align}
    \rho = \dfrac{1}{2^Q}\sum_{j_{Q-1}=0}^3\ldots \sum_{j_{0}=0}^3 r_{j_{Q-1},\ldots,j_0} \sigma_{j_{Q-1}}\otimes\ldots\otimes \sigma_{j_{0}}, 
\end{align}
where $\sigma_0 = I, \sigma_1=X, \sigma_2=Y,$ and $\sigma_3=Z$ are the Pauli matrices.
Therefore, the vector $(r_{j_{Q-1},\ldots,j_0})_{j_{Q-1}=0,\ldots,j_0=0}^{j_{Q-1}=3,\ldots,j_0=3} \in \bbR^{4^Q}$ can be considered as the multiqubit Bloch vector associated with $\rho$.
The condition $\tr[\rho]=1$ implies that $r_{0,...,0}=1$.
Therefore, we can represent $\rho$ with the vector form as 
\begin{align}\label{eqn:rho:vector}
\ket{\rho}\rangle =\dfrac{1}{2^Q}\begin{pmatrix} 1 \\ \br \end{pmatrix}. 
\end{align}
We can verify that $|\br| \leq \sqrt{2^Q - 1}$ and the equality occurs if and only if $\rho = \ket{\psi}\bra{\psi}$ with $\ket{\psi}$ is a pure $Q$-qubit state.
The $i$th element of $\begin{pmatrix} 1 \\ \br \end{pmatrix}$ is $\tr[P_i\rho]$, where $P_i=\sigma_{j_{Q-1}}\otimes\ldots\otimes \sigma_{j_{0}}$ is the $i$th Pauli string.

In general, a quantum channel $\cE$ acting on a density matrix $\rho$ can be written as applying a matrix operator $\hat{E}$ to the vector form of $\rho$ as
\begin{align}\label{eqn:ptm:channel}
\ket{\cE(\rho)}\rangle = \hat{E} \ket{\rho}\rangle.    
\end{align}
Here, $\hat{E}$ is the Pauli transfer matrix (PTM) representation  of the quantum channel $\cE$, which is represented as
\begin{align}\label{eqn:ptm:operator}
    \hat{E} = \begin{pmatrix} 1 & \boldsymbol{0}^\top \\ \boldsymbol{b} & W \end{pmatrix},
\end{align}
where $\boldsymbol{0}=(0,0,\ldots,0)\in \bbR^{4^Q-1}$, $\boldsymbol{b} \in \bbR^{4^Q-1}$ and $W \in \bbR^{(4^Q-1)\times (4^Q-1)}$.
Note that, if $\cE$ is a unitary channel then $\bb=\boldsymbol{0}$.

With this PTM representation of the quantum channel, Eq.~\eqref{eqn:ptm:channel} can be rewritten as
\begin{align}
    \br^{\prime} = \bb + W\br,
\end{align}
where $\ket{\cE(\rho)}\rangle =\dfrac{1}{2^Q}\begin{pmatrix} 1 \\ \br^{\prime}  \end{pmatrix}$.

We formulate the transfer learning minimax risk in terms of PTM representation.
We define the matrix $W$ in Eq.~\eqref{eqn:ptm:operator} corresponding to unitary matrices $V_m, V_M$, and $U(\bth_M)$ as $W_m, W_M$, and $W(\bth_M)$, respectively. 
We also define the vector $\br$ in Eq.~\eqref{eqn:rho:vector} corresponding to quantum states $\ket{\psi^{(M)}_j}$ as $\br^{(M)}_j$, and rewrite the PTM representation of the quantum channel $\cE(\cdot, \epsilon^{(M)}_j)$ as
\begin{align}
    \hat{\cE}(\cdot, \epsilon^{(M)}_j) = \begin{pmatrix} 1 & \boldsymbol{0}^\top \\ \boldsymbol{b}(\epsilon^{(M)}_j) & W( \epsilon^{(M)}_j) \end{pmatrix}.
\end{align}
The loss function in Eq.~\eqref{eqn:loss:transfer} can be expressed as
\begin{widetext}
\begin{align}
    \ell_j(\bth_M) =  \dfrac{1}{2^{Q+1}}\left\|  W(\bth_M)\br^{(M)}_j - \left( W(\epsilon^{(M)}_j)W_M \br^{(M)}_j +  \boldsymbol{b}(\epsilon^{(M)}_j) \right) \right\|^2 = \dfrac{1}{2^{Q+1}}\left\| \left( W(\bth_M) - W(\epsilon^{(M)}_j)W_M \right)\br^{(M)}_j -  \boldsymbol{b}(\epsilon^{(M)}_j) \right\|^2
\end{align}    
\end{widetext}



Therefore, we can adapt the minimax framework to the linear regression setting, similar to approaches in the classical context~\cite{kalan:2020:minimax}. It is essential to consider the requirements for $\br$ to ensure it can represent a physical state and to specify the representations of the noise channel $\cE$. For instance, if we only consider the unitary noise channel, then $\boldsymbol{b}(\epsilon^{(M)}_j) = \boldsymbol{0}$. We leave this intriguing aspect for future investigation.

\section{Formulation of the loss function for data-based Q-CurL}\label{sec:loss:func}

\begin{figure}
		\includegraphics[width=8.7cm]{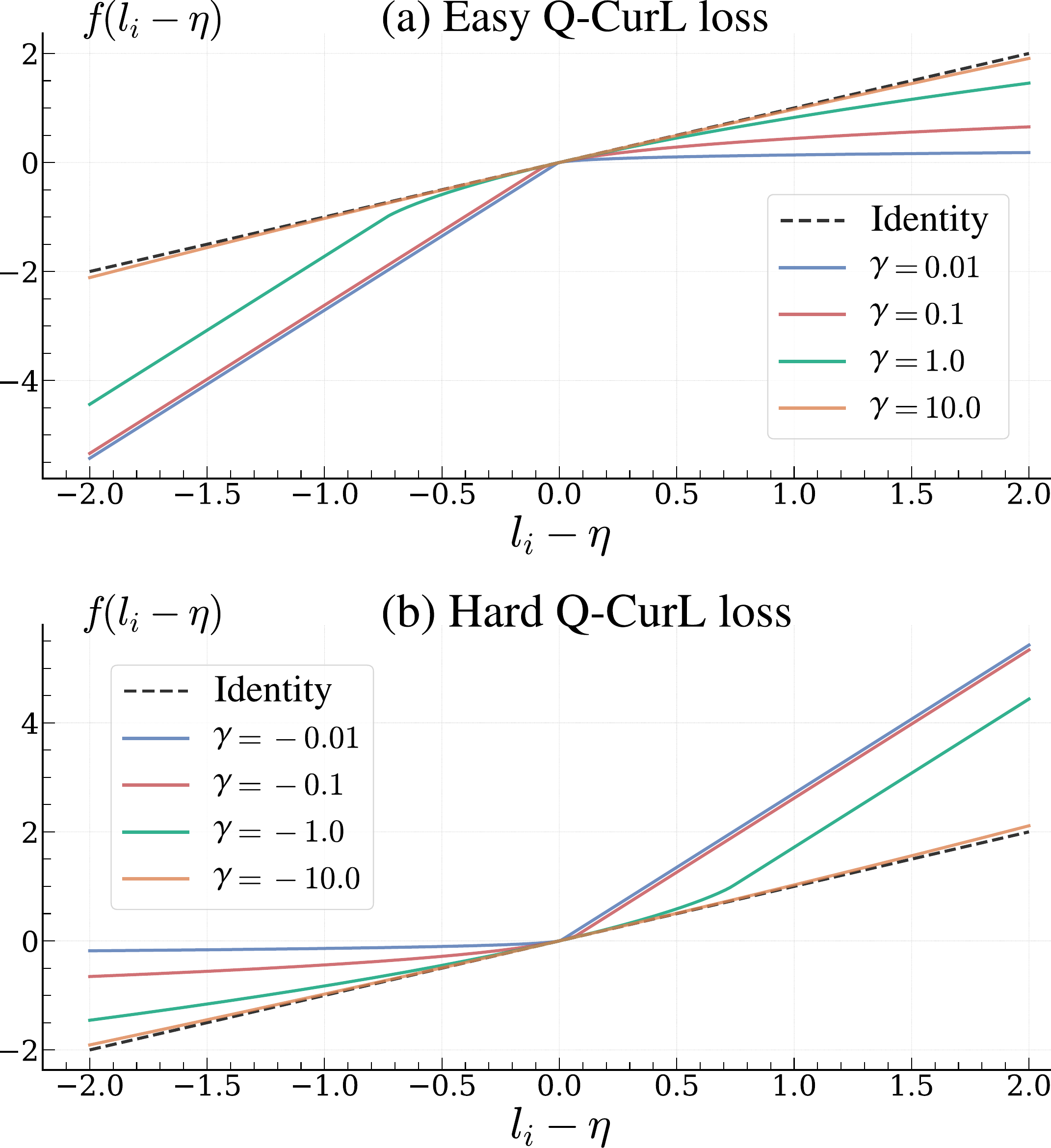}
		\protect\caption{Illustration of the function $f(l_i - \eta)$ as defined in Eq.~\eqref{eqn:single:loss} for different values of $\gamma$. The \textit{easy Q-CurL} scenario ($\gamma > 0$) emphasizes small losses, while the \textit{hard Q-CurL} scenario ($\gamma < 0)$ emphasizes large losses.
		\label{fig:curr:loss}}
\end{figure}

In data-based Q-CurL, we train with the loss 
\begin{align}
\hat{R}(h, \bw) = \dfrac{1}{N}\sum_{i=1}^N \left( ( \ell_i - \eta )e^{w_i} + \gamma w^2_i \right),
\end{align} and the procedure $\textup{min}_{\bth}\textup{min}_{\bw}\hat{R}(h, \bw)$ mentioned in the main text.
Here, $\textup{min}_{\bw}\hat{R}(h, \bw)$ is decomposed at each loss $\ell_i$ and solved without quantum resources as 
\begin{align}~\label{eqn:loss:min}
    w_i = \textup{argmin}_w(l_i-\eta)e^{w} + \gamma w^2.
\end{align} 

Let $a_i=\dfrac{l_i-\eta}{\gamma}$, we can reduce Eq.~\eqref{eqn:loss:min} into the following form:
$w_i = \textup{argmin}_w g(w),$
with $g(w)= a_ie^{w} + w^2$ is the function of the scalar variable $w$.

To solve the minimization $w_i = \textup{argmin}_w g(w)$, we consider zero point of the derivative of $g(w)$ as
\begin{align} \label{eqn:lambert}
    \dfrac{dg}{dw} = a_i e^w + 2w = 0 \Longleftrightarrow (-w)e^{-w} = \dfrac{a_i}{2}.
\end{align}
Equation~\eqref{eqn:lambert} yields a solution $w = -W(\dfrac{a_i}{2})$ only for $\dfrac{a_i}{2} \geq -\dfrac{1}{e}$. Here, $W(z)$ defined for $z\geq -\dfrac{1}{e}$ is called principal branch of Lambert W function that satisfies $W(z)e^{W(z)}=z$.
Since the principal branch of the Lambert W function is monotonically increasing, we set the weight $w_i =  -W\left(\textup{max}(-\dfrac{1}{e}, \dfrac{a_i}{2})\right) = -W\left(\textup{max}(-\dfrac{1}{e}, \dfrac{l_i-\eta}{2\gamma})\right)$.

Let $z_i = \textup{max}(-\dfrac{1}{e}, \dfrac{l_i - \eta}{2\gamma})$ then $e^{w_i} = -\dfrac{w_i}{z_i} = \dfrac{W(z_i)}{z_i}$, the modified loss of $l_i$ becomes
\begin{align}\label{eqn:single:loss}
    f(l_i - \eta) = (l_i-\eta)e^{w_i} + \gamma w_i^2 = (l_i-\eta)\dfrac{W(z_i)}{z_i}  + \gamma W^2(z_i).
\end{align}

We use the mpmath~\cite{mpmath} library to implement the Lambert W function and then plot the function $f(l_i - \eta)$ with different values of $\gamma$ in Fig.~\ref{fig:curr:loss}.

First, when $|\gamma|$ is sufficient large, $f(l_i-\mu) \approx l_i - \mu$. This approximation can be easily verified from Eq.~\eqref{eqn:single:loss} as $w_i\to 0$ when $\dfrac{l_i - \mu}{2\gamma} \to 0$.
For other values of $\gamma$, the sign of $\gamma$ determines whether the optimization process emphasizes easy samples ($l_i < \eta$) or hard samples ($l_i > \eta$). Specifically, if $\gamma > 0$, the slope of $f(l_i - \mu)$ is bigger for small losses ($l_i < \eta$) and smaller than the slope of the identity function for large losses $l_i > \eta$. Thus, the optimization process should prioritize emphasizing small losses.
Conversely, if $\gamma < 0$, the slope of $f(l_i - \mu)$ is smaller for small losses $l_i < \eta$ and larger than the slope of the identity function for large losses $l_i > \eta$. Thus, the optimization process should prioritize emphasizing large losses.
We define these two scenarios as \textit{easy Q-CurL} and \textit{hard Q-CurL}, as depicted in Fig.~\ref{fig:curr:loss}(a) and Fig.~\ref{fig:curr:loss}(b), respectively.

\bibliography{main.bib}

\providecommand{\noopsort}[1]{}\providecommand{\singleletter}[1]{#1}%
\begin{thebibliography}{52}%
\makeatletter
\providecommand \@ifxundefined [1]{%
 \@ifx{#1\undefined}
}%
\providecommand \@ifnum [1]{%
 \ifnum #1\expandafter \@firstoftwo
 \else \expandafter \@secondoftwo
 \fi
}%
\providecommand \@ifx [1]{%
 \ifx #1\expandafter \@firstoftwo
 \else \expandafter \@secondoftwo
 \fi
}%
\providecommand \natexlab [1]{#1}%
\providecommand \enquote  [1]{``#1''}%
\providecommand \bibnamefont  [1]{#1}%
\providecommand \bibfnamefont [1]{#1}%
\providecommand \citenamefont [1]{#1}%
\providecommand \href@noop [0]{\@secondoftwo}%
\providecommand \href [0]{\begingroup \@sanitize@url \@href}%
\providecommand \@href[1]{\@@startlink{#1}\@@href}%
\providecommand \@@href[1]{\endgroup#1\@@endlink}%
\providecommand \@sanitize@url [0]{\catcode `\\12\catcode `\$12\catcode `\&12\catcode `\#12\catcode `\^12\catcode `\_12\catcode `\%12\relax}%
\providecommand \@@startlink[1]{}%
\providecommand \@@endlink[0]{}%
\providecommand \url  [0]{\begingroup\@sanitize@url \@url }%
\providecommand \@url [1]{\endgroup\@href {#1}{\urlprefix }}%
\providecommand \urlprefix  [0]{URL }%
\providecommand \Eprint [0]{\href }%
\providecommand \doibase [0]{https://doi.org/}%
\providecommand \selectlanguage [0]{\@gobble}%
\providecommand \bibinfo  [0]{\@secondoftwo}%
\providecommand \bibfield  [0]{\@secondoftwo}%
\providecommand \translation [1]{[#1]}%
\providecommand \BibitemOpen [0]{}%
\providecommand \bibitemStop [0]{}%
\providecommand \bibitemNoStop [0]{.\EOS\space}%
\providecommand \EOS [0]{\spacefactor3000\relax}%
\providecommand \BibitemShut  [1]{\csname bibitem#1\endcsname}%
\let\auto@bib@innerbib\@empty
\bibitem [{\citenamefont {Biamonte}\ \emph {et~al.}(2017)\citenamefont {Biamonte}, \citenamefont {Wittek}, \citenamefont {Pancotti}, \citenamefont {Rebentrost}, \citenamefont {Wiebe},\ and\ \citenamefont {Lloyd}}]{biamonte:2017:QML}%
  \BibitemOpen
  \bibfield  {author} {\bibinfo {author} {\bibfnamefont {J.}~\bibnamefont {Biamonte}}, \bibinfo {author} {\bibfnamefont {P.}~\bibnamefont {Wittek}}, \bibinfo {author} {\bibfnamefont {N.}~\bibnamefont {Pancotti}}, \bibinfo {author} {\bibfnamefont {P.}~\bibnamefont {Rebentrost}}, \bibinfo {author} {\bibfnamefont {N.}~\bibnamefont {Wiebe}},\ and\ \bibinfo {author} {\bibfnamefont {S.}~\bibnamefont {Lloyd}},\ }\bibfield  {title} {\bibinfo {title} {Quantum machine learning},\ }\href {https://doi.org/10.1038/nature23474} {\bibfield  {journal} {\bibinfo  {journal} {Nature}\ }\textbf {\bibinfo {volume} {549}},\ \bibinfo {pages} {195} (\bibinfo {year} {2017})}\BibitemShut {NoStop}%
\bibitem [{\citenamefont {Schuld}\ and\ \citenamefont {Petruccione}(2021)}]{schuld:2021:qmlbook}%
  \BibitemOpen
  \bibfield  {author} {\bibinfo {author} {\bibfnamefont {M.}~\bibnamefont {Schuld}}\ and\ \bibinfo {author} {\bibfnamefont {F.}~\bibnamefont {Petruccione}},\ }\href {https://doi.org/10.1007/978-3-030-83098-4} {\emph {\bibinfo {title} {Machine Learning with Quantum Computers}}}\ (\bibinfo  {publisher} {Springer International Publishing},\ \bibinfo {year} {2021})\BibitemShut {NoStop}%
\bibitem [{\citenamefont {Harrow}\ \emph {et~al.}(2009)\citenamefont {Harrow}, \citenamefont {Hassidim},\ and\ \citenamefont {Lloyd}}]{HHL:2009:prl}%
  \BibitemOpen
  \bibfield  {author} {\bibinfo {author} {\bibfnamefont {A.~W.}\ \bibnamefont {Harrow}}, \bibinfo {author} {\bibfnamefont {A.}~\bibnamefont {Hassidim}},\ and\ \bibinfo {author} {\bibfnamefont {S.}~\bibnamefont {Lloyd}},\ }\bibfield  {title} {\bibinfo {title} {Quantum algorithm for linear systems of equations},\ }\href {https://doi.org/10.1103/PhysRevLett.103.150502} {\bibfield  {journal} {\bibinfo  {journal} {Phys. Rev. Lett.}\ }\textbf {\bibinfo {volume} {103}},\ \bibinfo {pages} {150502} (\bibinfo {year} {2009})}\BibitemShut {NoStop}%
\bibitem [{\citenamefont {Aaronson}(2015)}]{scott:2015:natphys}%
  \BibitemOpen
  \bibfield  {author} {\bibinfo {author} {\bibfnamefont {S.}~\bibnamefont {Aaronson}},\ }\bibfield  {title} {\bibinfo {title} {Read the fine print},\ }\href {https://doi.org/10.1038/nphys3272} {\bibfield  {journal} {\bibinfo  {journal} {Nat. Rev. Phys.}\ }\textbf {\bibinfo {volume} {11}},\ \bibinfo {pages} {291–293} (\bibinfo {year} {2015})}\BibitemShut {NoStop}%
\bibitem [{\citenamefont {Tang}(2022)}]{tang:2022:natphys}%
  \BibitemOpen
  \bibfield  {author} {\bibinfo {author} {\bibfnamefont {E.}~\bibnamefont {Tang}},\ }\bibfield  {title} {\bibinfo {title} {Dequantizing algorithms to understand quantum advantage in machine learning},\ }\href {https://doi.org/10.1038/s42254-022-00511-w} {\bibfield  {journal} {\bibinfo  {journal} {Nat. Rev. Phys.}\ }\textbf {\bibinfo {volume} {4}},\ \bibinfo {pages} {692–693} (\bibinfo {year} {2022})}\BibitemShut {NoStop}%
\bibitem [{\citenamefont {Mitarai}\ \emph {et~al.}(2018)\citenamefont {Mitarai}, \citenamefont {Negoro}, \citenamefont {Kitagawa},\ and\ \citenamefont {Fujii}}]{mitarai:2018:circuit}%
  \BibitemOpen
  \bibfield  {author} {\bibinfo {author} {\bibfnamefont {K.}~\bibnamefont {Mitarai}}, \bibinfo {author} {\bibfnamefont {M.}~\bibnamefont {Negoro}}, \bibinfo {author} {\bibfnamefont {M.}~\bibnamefont {Kitagawa}},\ and\ \bibinfo {author} {\bibfnamefont {K.}~\bibnamefont {Fujii}},\ }\bibfield  {title} {\bibinfo {title} {Quantum circuit learning},\ }\href {https://doi.org/10.1103/PhysRevA.98.032309} {\bibfield  {journal} {\bibinfo  {journal} {Phys. Rev. A}\ }\textbf {\bibinfo {volume} {98}},\ \bibinfo {pages} {032309} (\bibinfo {year} {2018})}\BibitemShut {NoStop}%
\bibitem [{\citenamefont {Schuld}\ \emph {et~al.}(2019)\citenamefont {Schuld}, \citenamefont {Bergholm}, \citenamefont {Gogolin}, \citenamefont {Izaac},\ and\ \citenamefont {Killoran}}]{schuld:2019:shift}%
  \BibitemOpen
  \bibfield  {author} {\bibinfo {author} {\bibfnamefont {M.}~\bibnamefont {Schuld}}, \bibinfo {author} {\bibfnamefont {V.}~\bibnamefont {Bergholm}}, \bibinfo {author} {\bibfnamefont {C.}~\bibnamefont {Gogolin}}, \bibinfo {author} {\bibfnamefont {J.}~\bibnamefont {Izaac}},\ and\ \bibinfo {author} {\bibfnamefont {N.}~\bibnamefont {Killoran}},\ }\bibfield  {title} {\bibinfo {title} {Evaluating analytic gradients on quantum hardware},\ }\href {https://doi.org/10.1103/PhysRevA.99.032331} {\bibfield  {journal} {\bibinfo  {journal} {Phys. Rev. A}\ }\textbf {\bibinfo {volume} {99}},\ \bibinfo {pages} {032331} (\bibinfo {year} {2019})}\BibitemShut {NoStop}%
\bibitem [{\citenamefont {Cerezo}\ \emph {et~al.}(2021)\citenamefont {Cerezo}, \citenamefont {Arrasmith}, \citenamefont {Babbush}, \citenamefont {Benjamin}, \citenamefont {Endo}, \citenamefont {Fujii}, \citenamefont {{McClean}}, \citenamefont {Mitarai}, \citenamefont {Yuan}, \citenamefont {Cincio},\ and\ \citenamefont {Coles}}]{cerezo:2021:variational:nature}%
  \BibitemOpen
  \bibfield  {author} {\bibinfo {author} {\bibfnamefont {M.}~\bibnamefont {Cerezo}}, \bibinfo {author} {\bibfnamefont {A.}~\bibnamefont {Arrasmith}}, \bibinfo {author} {\bibfnamefont {R.}~\bibnamefont {Babbush}}, \bibinfo {author} {\bibfnamefont {S.~C.}\ \bibnamefont {Benjamin}}, \bibinfo {author} {\bibfnamefont {S.}~\bibnamefont {Endo}}, \bibinfo {author} {\bibfnamefont {K.}~\bibnamefont {Fujii}}, \bibinfo {author} {\bibfnamefont {J.~R.}\ \bibnamefont {{McClean}}}, \bibinfo {author} {\bibfnamefont {K.}~\bibnamefont {Mitarai}}, \bibinfo {author} {\bibfnamefont {X.}~\bibnamefont {Yuan}}, \bibinfo {author} {\bibfnamefont {L.}~\bibnamefont {Cincio}},\ and\ \bibinfo {author} {\bibfnamefont {P.~J.}\ \bibnamefont {Coles}},\ }\bibfield  {title} {\bibinfo {title} {Variational quantum algorithms},\ }\href {https://doi.org/10.1038/s42254-021-00348-9} {\bibfield  {journal} {\bibinfo  {journal} {Nat. Rev. Phys.}\ }\textbf {\bibinfo {volume} {3}},\ \bibinfo {pages} {625} (\bibinfo {year} {2021})}\BibitemShut {NoStop}%
\bibitem [{\citenamefont {Havl{\'{\i}}{\v{c}}ek}\ \emph {et~al.}(2019)\citenamefont {Havl{\'{\i}}{\v{c}}ek}, \citenamefont {C{\'{o}}rcoles}, \citenamefont {Temme}, \citenamefont {Harrow}, \citenamefont {Kandala}, \citenamefont {Chow},\ and\ \citenamefont {Gambetta}}]{halvlicek:2019:supervised}%
  \BibitemOpen
  \bibfield  {author} {\bibinfo {author} {\bibfnamefont {V.}~\bibnamefont {Havl{\'{\i}}{\v{c}}ek}}, \bibinfo {author} {\bibfnamefont {A.~D.}\ \bibnamefont {C{\'{o}}rcoles}}, \bibinfo {author} {\bibfnamefont {K.}~\bibnamefont {Temme}}, \bibinfo {author} {\bibfnamefont {A.~W.}\ \bibnamefont {Harrow}}, \bibinfo {author} {\bibfnamefont {A.}~\bibnamefont {Kandala}}, \bibinfo {author} {\bibfnamefont {J.~M.}\ \bibnamefont {Chow}},\ and\ \bibinfo {author} {\bibfnamefont {J.~M.}\ \bibnamefont {Gambetta}},\ }\bibfield  {title} {\bibinfo {title} {Supervised learning with quantum-enhanced feature spaces},\ }\href {https://doi.org/10.1038/s41586-019-0980-2} {\bibfield  {journal} {\bibinfo  {journal} {Nature}\ }\textbf {\bibinfo {volume} {567}},\ \bibinfo {pages} {209} (\bibinfo {year} {2019})}\BibitemShut {NoStop}%
\bibitem [{\citenamefont {Schuld}\ and\ \citenamefont {Killoran}(2019)}]{schuld:2019:feature}%
  \BibitemOpen
  \bibfield  {author} {\bibinfo {author} {\bibfnamefont {M.}~\bibnamefont {Schuld}}\ and\ \bibinfo {author} {\bibfnamefont {N.}~\bibnamefont {Killoran}},\ }\bibfield  {title} {\bibinfo {title} {Quantum machine learning in feature \text{Hilbert} spaces},\ }\href {https://doi.org/10.1103/PhysRevLett.122.040504} {\bibfield  {journal} {\bibinfo  {journal} {Phys. Rev. Lett.}\ }\textbf {\bibinfo {volume} {122}},\ \bibinfo {pages} {040504} (\bibinfo {year} {2019})}\BibitemShut {NoStop}%
\bibitem [{\citenamefont {Fujii}\ and\ \citenamefont {Nakajima}(2017)}]{fujii:2017:qrc}%
  \BibitemOpen
  \bibfield  {author} {\bibinfo {author} {\bibfnamefont {K.}~\bibnamefont {Fujii}}\ and\ \bibinfo {author} {\bibfnamefont {K.}~\bibnamefont {Nakajima}},\ }\bibfield  {title} {\bibinfo {title} {Harnessing disordered-ensemble quantum dynamics for machine learning},\ }\href {https://doi.org/10.1103/PhysRevApplied.8.024030} {\bibfield  {journal} {\bibinfo  {journal} {Phys. Rev. Applied}\ }\textbf {\bibinfo {volume} {8}},\ \bibinfo {pages} {024030} (\bibinfo {year} {2017})}\BibitemShut {NoStop}%
\bibitem [{\citenamefont {Liu}\ \emph {et~al.}(2021)\citenamefont {Liu}, \citenamefont {Arunachalam},\ and\ \citenamefont {Temme}}]{liu:2020:rigorous}%
  \BibitemOpen
  \bibfield  {author} {\bibinfo {author} {\bibfnamefont {Y.}~\bibnamefont {Liu}}, \bibinfo {author} {\bibfnamefont {S.}~\bibnamefont {Arunachalam}},\ and\ \bibinfo {author} {\bibfnamefont {K.}~\bibnamefont {Temme}},\ }\bibfield  {title} {\bibinfo {title} {A rigorous and robust quantum speed-up in supervised machine learning},\ }\href {https://doi.org/10.1038/s41567-021-01287-z} {\bibfield  {journal} {\bibinfo  {journal} {Nat. Phys.}\ } (\bibinfo {year} {2021})}\BibitemShut {NoStop}%
\bibitem [{\citenamefont {Goto}\ \emph {et~al.}(2021)\citenamefont {Goto}, \citenamefont {Tran},\ and\ \citenamefont {Nakajima}}]{tran:2021:prl:uap}%
  \BibitemOpen
  \bibfield  {author} {\bibinfo {author} {\bibfnamefont {T.}~\bibnamefont {Goto}}, \bibinfo {author} {\bibfnamefont {Q.~H.}\ \bibnamefont {Tran}},\ and\ \bibinfo {author} {\bibfnamefont {K.}~\bibnamefont {Nakajima}},\ }\bibfield  {title} {\bibinfo {title} {Universal approximation property of quantum machine learning models in quantum-enhanced feature spaces},\ }\href {https://doi.org/10.1103/PhysRevLett.127.090506} {\bibfield  {journal} {\bibinfo  {journal} {Phys. Rev. Lett.}\ }\textbf {\bibinfo {volume} {127}},\ \bibinfo {pages} {090506} (\bibinfo {year} {2021})}\BibitemShut {NoStop}%
\bibitem [{\citenamefont {Gao}\ \emph {et~al.}(2022)\citenamefont {Gao}, \citenamefont {Anschuetz}, \citenamefont {Wang}, \citenamefont {Cirac},\ and\ \citenamefont {Lukin}}]{gao:2022:prx:corr}%
  \BibitemOpen
  \bibfield  {author} {\bibinfo {author} {\bibfnamefont {X.}~\bibnamefont {Gao}}, \bibinfo {author} {\bibfnamefont {E.~R.}\ \bibnamefont {Anschuetz}}, \bibinfo {author} {\bibfnamefont {S.-T.}\ \bibnamefont {Wang}}, \bibinfo {author} {\bibfnamefont {J.~I.}\ \bibnamefont {Cirac}},\ and\ \bibinfo {author} {\bibfnamefont {M.~D.}\ \bibnamefont {Lukin}},\ }\bibfield  {title} {\bibinfo {title} {Enhancing generative models via quantum correlations},\ }\href {https://doi.org/10.1103/PhysRevX.12.021037} {\bibfield  {journal} {\bibinfo  {journal} {Phys. Rev. X}\ }\textbf {\bibinfo {volume} {12}},\ \bibinfo {pages} {021037} (\bibinfo {year} {2022})}\BibitemShut {NoStop}%
\bibitem [{\citenamefont {Tran}\ and\ \citenamefont {Nakajima}(2020)}]{tran:2020:higherorder}%
  \BibitemOpen
  \bibfield  {author} {\bibinfo {author} {\bibfnamefont {Q.~H.}\ \bibnamefont {Tran}}\ and\ \bibinfo {author} {\bibfnamefont {K.}~\bibnamefont {Nakajima}},\ }\bibfield  {title} {\bibinfo {title} {Higher-order quantum reservoir computing},\ }\bibfield  {journal} {\bibinfo  {journal} {arXiv}\ }\href {https://doi.org/10.48550/arXiv.2006.08999} {10.48550/arXiv.2006.08999} (\bibinfo {year} {2020})\BibitemShut {NoStop}%
\bibitem [{\citenamefont {Schuld}\ and\ \citenamefont {Killoran}(2022)}]{schuld:2022:qadv}%
  \BibitemOpen
  \bibfield  {author} {\bibinfo {author} {\bibfnamefont {M.}~\bibnamefont {Schuld}}\ and\ \bibinfo {author} {\bibfnamefont {N.}~\bibnamefont {Killoran}},\ }\bibfield  {title} {\bibinfo {title} {Is quantum advantage the right goal for quantum machine learning?},\ }\href {https://doi.org/10.1103/PRXQuantum.3.030101} {\bibfield  {journal} {\bibinfo  {journal} {PRX Quantum}\ }\textbf {\bibinfo {volume} {3}},\ \bibinfo {pages} {030101} (\bibinfo {year} {2022})}\BibitemShut {NoStop}%
\bibitem [{\citenamefont {Editorial}(2023)}]{editorial:2023:qmladv:nature}%
  \BibitemOpen
  \bibfield  {author} {\bibinfo {author} {\bibnamefont {Editorial}},\ }\bibfield  {title} {\bibinfo {title} {Seeking a quantum advantage for machine learning},\ }\href {https://doi.org/10.1038/s42256-023-00710-9} {\bibfield  {journal} {\bibinfo  {journal} {Nat. Mach. Intell.}\ }\textbf {\bibinfo {volume} {5}},\ \bibinfo {pages} {813–813} (\bibinfo {year} {2023})}\BibitemShut {NoStop}%
\bibitem [{\citenamefont {Cong}\ \emph {et~al.}(2019)\citenamefont {Cong}, \citenamefont {Choi},\ and\ \citenamefont {Lukin}}]{cong:2019:QCNN}%
  \BibitemOpen
  \bibfield  {author} {\bibinfo {author} {\bibfnamefont {I.}~\bibnamefont {Cong}}, \bibinfo {author} {\bibfnamefont {S.}~\bibnamefont {Choi}},\ and\ \bibinfo {author} {\bibfnamefont {M.~D.}\ \bibnamefont {Lukin}},\ }\bibfield  {title} {\bibinfo {title} {{Quantum convolutional neural networks}},\ }\href {https://doi.org/10.1038/s41567-019-0648-8} {\bibfield  {journal} {\bibinfo  {journal} {Nat. Phys.}\ }\textbf {\bibinfo {volume} {15}},\ \bibinfo {pages} {1273} (\bibinfo {year} {2019})}\BibitemShut {NoStop}%
\bibitem [{\citenamefont {Perrier}\ \emph {et~al.}(2022)\citenamefont {Perrier}, \citenamefont {Youssry},\ and\ \citenamefont {Ferrie}}]{Perrier:2022:qdata}%
  \BibitemOpen
  \bibfield  {author} {\bibinfo {author} {\bibfnamefont {E.}~\bibnamefont {Perrier}}, \bibinfo {author} {\bibfnamefont {A.}~\bibnamefont {Youssry}},\ and\ \bibinfo {author} {\bibfnamefont {C.}~\bibnamefont {Ferrie}},\ }\bibfield  {title} {\bibinfo {title} {Qdataset, quantum datasets for machine learning},\ }\href {https://doi.org/10.1038/s41597-022-01639-1} {\bibfield  {journal} {\bibinfo  {journal} {Sci. Data}\ }\textbf {\bibinfo {volume} {9}},\ \bibinfo {pages} {582} (\bibinfo {year} {2022})}\BibitemShut {NoStop}%
\bibitem [{\citenamefont {Haug}\ and\ \citenamefont {Kim}(2024)}]{haug:2023:generalization}%
  \BibitemOpen
  \bibfield  {author} {\bibinfo {author} {\bibfnamefont {T.}~\bibnamefont {Haug}}\ and\ \bibinfo {author} {\bibfnamefont {M.~S.}\ \bibnamefont {Kim}},\ }\bibfield  {title} {\bibinfo {title} {Generalization of quantum machine learning models using quantum fisher information metric},\ }\href {https://doi.org/10.1103/PhysRevLett.133.050603} {\bibfield  {journal} {\bibinfo  {journal} {Phys. Rev. Lett.}\ }\textbf {\bibinfo {volume} {133}},\ \bibinfo {pages} {050603} (\bibinfo {year} {2024})}\BibitemShut {NoStop}%
\bibitem [{\citenamefont {Chinzei}\ \emph {et~al.}(2024)\citenamefont {Chinzei}, \citenamefont {Tran}, \citenamefont {Maruyama}, \citenamefont {Oshima},\ and\ \citenamefont {Sato}}]{chinzei:2024:spQCNN}%
  \BibitemOpen
  \bibfield  {author} {\bibinfo {author} {\bibfnamefont {K.}~\bibnamefont {Chinzei}}, \bibinfo {author} {\bibfnamefont {Q.~H.}\ \bibnamefont {Tran}}, \bibinfo {author} {\bibfnamefont {K.}~\bibnamefont {Maruyama}}, \bibinfo {author} {\bibfnamefont {H.}~\bibnamefont {Oshima}},\ and\ \bibinfo {author} {\bibfnamefont {S.}~\bibnamefont {Sato}},\ }\bibfield  {title} {\bibinfo {title} {Splitting and parallelizing of quantum convolutional neural networks for learning translationally symmetric data},\ }\href {https://doi.org/10.1103/PhysRevResearch.6.023042} {\bibfield  {journal} {\bibinfo  {journal} {Phys. Rev. Res.}\ }\textbf {\bibinfo {volume} {6}},\ \bibinfo {pages} {023042} (\bibinfo {year} {2024})}\BibitemShut {NoStop}%
\bibitem [{\citenamefont {Tran}\ \emph {et~al.}(2024)\citenamefont {Tran}, \citenamefont {Kikuchi},\ and\ \citenamefont {Oshima}}]{tran:2024:varQAE}%
  \BibitemOpen
  \bibfield  {author} {\bibinfo {author} {\bibfnamefont {Q.~H.}\ \bibnamefont {Tran}}, \bibinfo {author} {\bibfnamefont {S.}~\bibnamefont {Kikuchi}},\ and\ \bibinfo {author} {\bibfnamefont {H.}~\bibnamefont {Oshima}},\ }\bibfield  {title} {\bibinfo {title} {Variational denoising for variational quantum eigensolver},\ }\href {https://doi.org/10.1103/PhysRevResearch.6.023181} {\bibfield  {journal} {\bibinfo  {journal} {Phys. Rev. Res.}\ }\textbf {\bibinfo {volume} {6}},\ \bibinfo {pages} {023181} (\bibinfo {year} {2024})}\BibitemShut {NoStop}%
\bibitem [{\citenamefont {Bittel}\ and\ \citenamefont {Kliesch}(2021)}]{bittel:2021:prl:VQANP}%
  \BibitemOpen
  \bibfield  {author} {\bibinfo {author} {\bibfnamefont {L.}~\bibnamefont {Bittel}}\ and\ \bibinfo {author} {\bibfnamefont {M.}~\bibnamefont {Kliesch}},\ }\bibfield  {title} {\bibinfo {title} {Training variational quantum algorithms is {NP}-hard},\ }\href {https://doi.org/10.1103/PhysRevLett.127.120502} {\bibfield  {journal} {\bibinfo  {journal} {Phys. Rev. Lett.}\ }\textbf {\bibinfo {volume} {127}},\ \bibinfo {pages} {120502} (\bibinfo {year} {2021})}\BibitemShut {NoStop}%
\bibitem [{\citenamefont {Anschuetz}\ and\ \citenamefont {Kiani}(2022)}]{ansachuetz:2022:natcom:VQA}%
  \BibitemOpen
  \bibfield  {author} {\bibinfo {author} {\bibfnamefont {E.~R.}\ \bibnamefont {Anschuetz}}\ and\ \bibinfo {author} {\bibfnamefont {B.~T.}\ \bibnamefont {Kiani}},\ }\bibfield  {title} {\bibinfo {title} {{Quantum variational algorithms are swamped with traps}},\ }\href {https://doi.org/10.1038/s41467-022-35364-5} {\bibfield  {journal} {\bibinfo  {journal} {Nat. Commun.}\ }\textbf {\bibinfo {volume} {13}},\ \bibinfo {pages} {7760} (\bibinfo {year} {2022})}\BibitemShut {NoStop}%
\bibitem [{\citenamefont {McClean}\ \emph {et~al.}(2018)\citenamefont {McClean}, \citenamefont {Boixo}, \citenamefont {Smelyanskiy}, \citenamefont {Babbush},\ and\ \citenamefont {Neven}}]{clean:2018:natcom:barren}%
  \BibitemOpen
  \bibfield  {author} {\bibinfo {author} {\bibfnamefont {J.~R.}\ \bibnamefont {McClean}}, \bibinfo {author} {\bibfnamefont {S.}~\bibnamefont {Boixo}}, \bibinfo {author} {\bibfnamefont {V.~N.}\ \bibnamefont {Smelyanskiy}}, \bibinfo {author} {\bibfnamefont {R.}~\bibnamefont {Babbush}},\ and\ \bibinfo {author} {\bibfnamefont {H.}~\bibnamefont {Neven}},\ }\bibfield  {title} {\bibinfo {title} {{Barren plateaus in quantum neural network training landscapes}},\ }\href {https://doi.org/10.1038/s41467-018-07090-4} {\bibfield  {journal} {\bibinfo  {journal} {Nat. Commun.}\ }\textbf {\bibinfo {volume} {9}},\ \bibinfo {pages} {4812} (\bibinfo {year} {2018})}\BibitemShut {NoStop}%
\bibitem [{\citenamefont {Chollet}(2019)}]{chollet:2019:intelligence}%
  \BibitemOpen
  \bibfield  {author} {\bibinfo {author} {\bibfnamefont {F.}~\bibnamefont {Chollet}},\ }\bibfield  {title} {\bibinfo {title} {On the measure of intelligence},\ }\bibfield  {journal} {\bibinfo  {journal} {arXiv}\ }\href {https://doi.org/10.48550/arXiv.1911.01547} {10.48550/arXiv.1911.01547} (\bibinfo {year} {2019})\BibitemShut {NoStop}%
\bibitem [{\citenamefont {Bengio}\ \emph {et~al.}(2009)\citenamefont {Bengio}, \citenamefont {Louradour}, \citenamefont {Collobert},\ and\ \citenamefont {Weston}}]{bengio:2009:curr}%
  \BibitemOpen
  \bibfield  {author} {\bibinfo {author} {\bibfnamefont {Y.}~\bibnamefont {Bengio}}, \bibinfo {author} {\bibfnamefont {J.}~\bibnamefont {Louradour}}, \bibinfo {author} {\bibfnamefont {R.}~\bibnamefont {Collobert}},\ and\ \bibinfo {author} {\bibfnamefont {J.}~\bibnamefont {Weston}},\ }\bibfield  {title} {\bibinfo {title} {Curriculum learning},\ }\href {https://doi.org/10.1145/1553374.1553380} {\bibfield  {journal} {\bibinfo  {journal} {Proc. 26th Int. Conf. Mach. Learn.}\ }\bibinfo {series} {ICML'09},\ \bibinfo {pages} {41–48} (\bibinfo {year} {2009})}\BibitemShut {NoStop}%
\bibitem [{\citenamefont {Soviany}\ \emph {et~al.}(2022)\citenamefont {Soviany}, \citenamefont {Ionescu}, \citenamefont {Rota},\ and\ \citenamefont {Sebe}}]{soviany:2022:curr:survey}%
  \BibitemOpen
  \bibfield  {author} {\bibinfo {author} {\bibfnamefont {P.}~\bibnamefont {Soviany}}, \bibinfo {author} {\bibfnamefont {R.~T.}\ \bibnamefont {Ionescu}}, \bibinfo {author} {\bibfnamefont {P.}~\bibnamefont {Rota}},\ and\ \bibinfo {author} {\bibfnamefont {N.}~\bibnamefont {Sebe}},\ }\bibfield  {title} {\bibinfo {title} {Curriculum learning: A survey},\ }\href {https://doi.org/10.1007/s11263-022-01611-x} {\bibfield  {journal} {\bibinfo  {journal} {Int. J. Comput. Vision}\ }\textbf {\bibinfo {volume} {130}},\ \bibinfo {pages} {1526–1565} (\bibinfo {year} {2022})}\BibitemShut {NoStop}%
\bibitem [{\citenamefont {Zhang}\ \emph {et~al.}(2015)\citenamefont {Zhang}, \citenamefont {Meng}, \citenamefont {Li}, \citenamefont {Jiang}, \citenamefont {Zhao},\ and\ \citenamefont {Han}}]{Zhang:2015:diversity:curr}%
  \BibitemOpen
  \bibfield  {author} {\bibinfo {author} {\bibfnamefont {D.}~\bibnamefont {Zhang}}, \bibinfo {author} {\bibfnamefont {D.}~\bibnamefont {Meng}}, \bibinfo {author} {\bibfnamefont {C.}~\bibnamefont {Li}}, \bibinfo {author} {\bibfnamefont {L.}~\bibnamefont {Jiang}}, \bibinfo {author} {\bibfnamefont {Q.}~\bibnamefont {Zhao}},\ and\ \bibinfo {author} {\bibfnamefont {J.}~\bibnamefont {Han}},\ }\bibfield  {title} {\bibinfo {title} {A self-paced multiple-instance learning framework for co-saliency detection},\ }in\ \href {https://openaccess.thecvf.com/content_iccv_2015/papers/Zhang_A_Self-Paced_Multiple-Instance_ICCV_2015_paper.pdf} {\emph {\bibinfo {booktitle} {Proc. IEEE Int. Conf. Comput. Vis}}},\ \bibinfo {series and number} {ICCV'15}\ (\bibinfo {year} {2015})\BibitemShut {NoStop}%
\bibitem [{\citenamefont {Soviany}(2020)}]{soviany:2020:curriculum}%
  \BibitemOpen
  \bibfield  {author} {\bibinfo {author} {\bibfnamefont {P.}~\bibnamefont {Soviany}},\ }\bibfield  {title} {\bibinfo {title} {Curriculum learning with diversity for supervised computer vision tasks},\ }in\ \href {https://openreview.net/forum?id=WH27bUkkzj} {\emph {\bibinfo {booktitle} {4th Lifelong Machine Learning Workshop at ICML 2020}}}\ (\bibinfo {year} {2020})\BibitemShut {NoStop}%
\bibitem [{\citenamefont {Jiang}\ \emph {et~al.}(2015)\citenamefont {Jiang}, \citenamefont {Meng}, \citenamefont {Zhao}, \citenamefont {Shan},\ and\ \citenamefont {Hauptmann}}]{jiang:2015:self-paced}%
  \BibitemOpen
  \bibfield  {author} {\bibinfo {author} {\bibfnamefont {L.}~\bibnamefont {Jiang}}, \bibinfo {author} {\bibfnamefont {D.}~\bibnamefont {Meng}}, \bibinfo {author} {\bibfnamefont {Q.}~\bibnamefont {Zhao}}, \bibinfo {author} {\bibfnamefont {S.}~\bibnamefont {Shan}},\ and\ \bibinfo {author} {\bibfnamefont {A.}~\bibnamefont {Hauptmann}},\ }\bibfield  {title} {\bibinfo {title} {Self-paced curriculum learning},\ }in\ \href {https://doi.org/10.1609/aaai.v29i1.9608} {\emph {\bibinfo {booktitle} {Proc. 29th AAAI Conf. Artif. Intell.}}},\ \bibinfo {series and number} {\bibinfo {series} {AAAI'15}\ No.~\bibinfo {number} {1}}\ (\bibinfo {year} {2015})\BibitemShut {NoStop}%
\bibitem [{\citenamefont {Karras}\ \emph {et~al.}(2018)\citenamefont {Karras}, \citenamefont {Aila}, \citenamefont {Laine},\ and\ \citenamefont {Lehtinen}}]{karras:2018:progressive}%
  \BibitemOpen
  \bibfield  {author} {\bibinfo {author} {\bibfnamefont {T.}~\bibnamefont {Karras}}, \bibinfo {author} {\bibfnamefont {T.}~\bibnamefont {Aila}}, \bibinfo {author} {\bibfnamefont {S.}~\bibnamefont {Laine}},\ and\ \bibinfo {author} {\bibfnamefont {J.}~\bibnamefont {Lehtinen}},\ }\bibfield  {title} {\bibinfo {title} {Progressive growing of {GAN}s for improved quality, stability, and variation},\ }in\ \href {https://openreview.net/forum?id=Hk99zCeAb} {\emph {\bibinfo {booktitle} {Proc. Int. Conf. Learn. Represent.}}},\ \bibinfo {series and number} {ICLR'08}\ (\bibinfo {year} {2018})\BibitemShut {NoStop}%
\bibitem [{\citenamefont {Matiisen}\ \emph {et~al.}(2020)\citenamefont {Matiisen}, \citenamefont {Oliver}, \citenamefont {Cohen},\ and\ \citenamefont {Schulman}}]{matiisen:2020:teacher}%
  \BibitemOpen
  \bibfield  {author} {\bibinfo {author} {\bibfnamefont {T.}~\bibnamefont {Matiisen}}, \bibinfo {author} {\bibfnamefont {A.}~\bibnamefont {Oliver}}, \bibinfo {author} {\bibfnamefont {T.}~\bibnamefont {Cohen}},\ and\ \bibinfo {author} {\bibfnamefont {J.}~\bibnamefont {Schulman}},\ }\bibfield  {title} {\bibinfo {title} {Teacher–student curriculum learning},\ }\href {https://doi.org/10.1109/tnnls.2019.2934906} {\bibfield  {journal} {\bibinfo  {journal} {IEEE Trans. Neural Netw. Learn. Syst.}\ }\textbf {\bibinfo {volume} {31}},\ \bibinfo {pages} {3732–3740} (\bibinfo {year} {2020})}\BibitemShut {NoStop}%
\bibitem [{\citenamefont {Sinha}\ \emph {et~al.}(2020)\citenamefont {Sinha}, \citenamefont {Garg},\ and\ \citenamefont {Larochelle}}]{sinha:2020:implicit}%
  \BibitemOpen
  \bibfield  {author} {\bibinfo {author} {\bibfnamefont {S.}~\bibnamefont {Sinha}}, \bibinfo {author} {\bibfnamefont {A.}~\bibnamefont {Garg}},\ and\ \bibinfo {author} {\bibfnamefont {H.}~\bibnamefont {Larochelle}},\ }\bibfield  {title} {\bibinfo {title} {Curriculum by smoothing},\ }in\ \href {https://proceedings.neurips.cc/paper_files/paper/2020/file/f6a673f09493afcd8b129a0bcf1cd5bc-Paper.pdf} {\emph {\bibinfo {booktitle} {Adv. Neural Inf. Process. Syst.}}},\ Vol.~\bibinfo {volume} {33},\ \bibinfo {editor} {edited by\ \bibinfo {editor} {\bibfnamefont {H.}~\bibnamefont {Larochelle}}, \bibinfo {editor} {\bibfnamefont {M.}~\bibnamefont {Ranzato}}, \bibinfo {editor} {\bibfnamefont {R.}~\bibnamefont {Hadsell}}, \bibinfo {editor} {\bibfnamefont {M.}~\bibnamefont {Balcan}},\ and\ \bibinfo {editor} {\bibfnamefont {H.}~\bibnamefont {Lin}}}\ (\bibinfo  {publisher} {Curran Associates, Inc.},\ \bibinfo {year} {2020})\ pp.\ \bibinfo {pages} {21653--21664}\BibitemShut {NoStop}%
\bibitem [{\citenamefont {Saxena}\ \emph {et~al.}(2019)\citenamefont {Saxena}, \citenamefont {Tuzel},\ and\ \citenamefont {DeCoste}}]{saxena:2019:dataparm:NIPS}%
  \BibitemOpen
  \bibfield  {author} {\bibinfo {author} {\bibfnamefont {S.}~\bibnamefont {Saxena}}, \bibinfo {author} {\bibfnamefont {O.}~\bibnamefont {Tuzel}},\ and\ \bibinfo {author} {\bibfnamefont {D.}~\bibnamefont {DeCoste}},\ }\bibfield  {title} {\bibinfo {title} {Data parameters: A new family of parameters for learning a differentiable curriculum},\ }in\ \href {https://proceedings.neurips.cc/paper_files/paper/2019/file/926ffc0ca56636b9e73c565cf994ea5a-Paper.pdf} {\emph {\bibinfo {booktitle} {Adv. Neural Inf. Process. Syst.}}},\ Vol.~\bibinfo {volume} {32},\ \bibinfo {editor} {edited by\ \bibinfo {editor} {\bibfnamefont {H.}~\bibnamefont {Wallach}}, \bibinfo {editor} {\bibfnamefont {H.}~\bibnamefont {Larochelle}}, \bibinfo {editor} {\bibfnamefont {A.}~\bibnamefont {Beygelzimer}}, \bibinfo {editor} {\bibfnamefont {F.}~\bibnamefont {d\textquotesingle Alch\'{e}-Buc}}, \bibinfo {editor} {\bibfnamefont {E.}~\bibnamefont {Fox}},\ and\ \bibinfo {editor} {\bibfnamefont {R.}~\bibnamefont {Garnett}}}\ (\bibinfo  {publisher}
  {Curran Associates, Inc.},\ \bibinfo {year} {2019})\BibitemShut {NoStop}%
\bibitem [{\citenamefont {Castells}\ \emph {et~al.}(2020)\citenamefont {Castells}, \citenamefont {Weinzaepfel},\ and\ \citenamefont {Revaud}}]{superloss:2020:NIPS}%
  \BibitemOpen
  \bibfield  {author} {\bibinfo {author} {\bibfnamefont {T.}~\bibnamefont {Castells}}, \bibinfo {author} {\bibfnamefont {P.}~\bibnamefont {Weinzaepfel}},\ and\ \bibinfo {author} {\bibfnamefont {J.}~\bibnamefont {Revaud}},\ }\bibfield  {title} {\bibinfo {title} {Superloss: A generic loss for robust curriculum learning},\ }in\ \href {https://proceedings.neurips.cc/paper_files/paper/2020/file/2cfa8f9e50e0f510ede9d12338a5f564-Paper.pdf} {\emph {\bibinfo {booktitle} {Adv. Neural Inf. Process. Syst.}}},\ Vol.~\bibinfo {volume} {33},\ \bibinfo {editor} {edited by\ \bibinfo {editor} {\bibfnamefont {H.}~\bibnamefont {Larochelle}}, \bibinfo {editor} {\bibfnamefont {M.}~\bibnamefont {Ranzato}}, \bibinfo {editor} {\bibfnamefont {R.}~\bibnamefont {Hadsell}}, \bibinfo {editor} {\bibfnamefont {M.}~\bibnamefont {Balcan}},\ and\ \bibinfo {editor} {\bibfnamefont {H.}~\bibnamefont {Lin}}}\ (\bibinfo  {publisher} {Curran Associates, Inc.},\ \bibinfo {year} {2020})\ pp.\ \bibinfo {pages} {4308--4319}\BibitemShut {NoStop}%
\bibitem [{\citenamefont {Wang}\ \emph {et~al.}(2019)\citenamefont {Wang}, \citenamefont {Gan}, \citenamefont {Yang}, \citenamefont {Wu},\ and\ \citenamefont {Yan}}]{wang:2019:dynamic:curr}%
  \BibitemOpen
  \bibfield  {author} {\bibinfo {author} {\bibfnamefont {Y.}~\bibnamefont {Wang}}, \bibinfo {author} {\bibfnamefont {W.}~\bibnamefont {Gan}}, \bibinfo {author} {\bibfnamefont {J.}~\bibnamefont {Yang}}, \bibinfo {author} {\bibfnamefont {W.}~\bibnamefont {Wu}},\ and\ \bibinfo {author} {\bibfnamefont {J.}~\bibnamefont {Yan}},\ }\bibfield  {title} {\bibinfo {title} {Dynamic curriculum learning for imbalanced data classification},\ }in\ \href {https://openaccess.thecvf.com/content_ICCV_2019/papers/Wang_Dynamic_Curriculum_Learning_for_Imbalanced_Data_Classification_ICCV_2019_paper.pdf} {\emph {\bibinfo {booktitle} {Proc. IEEE Int. Conf. Comput. Vis.}}},\ \bibinfo {series and number} {ICCV'19}\ (\bibinfo {year} {2019})\ pp.\ \bibinfo {pages} {5017--5026}\BibitemShut {NoStop}%
\bibitem [{\citenamefont {Novotny}\ \emph {et~al.}(2018)\citenamefont {Novotny}, \citenamefont {Albanie}, \citenamefont {Larlus},\ and\ \citenamefont {Vedaldi}}]{novot:2018:CVPR}%
  \BibitemOpen
  \bibfield  {author} {\bibinfo {author} {\bibfnamefont {D.}~\bibnamefont {Novotny}}, \bibinfo {author} {\bibfnamefont {S.}~\bibnamefont {Albanie}}, \bibinfo {author} {\bibfnamefont {D.}~\bibnamefont {Larlus}},\ and\ \bibinfo {author} {\bibfnamefont {A.}~\bibnamefont {Vedaldi}},\ }\bibfield  {title} {\bibinfo {title} {Self-supervised learning of geometrically stable features through probabilistic introspection},\ }in\ \href {https://doi.org/10.1109/cvpr.2018.00383} {\emph {\bibinfo {booktitle} {Proc. IEEE/CVF Conf. Comput. Vis. Pattern Recognit.}}},\ \bibinfo {series and number} {CVPR'18}\ (\bibinfo  {publisher} {IEEE},\ \bibinfo {year} {2018})\BibitemShut {NoStop}%
\bibitem [{\citenamefont {Xu}\ \emph {et~al.}(2020)\citenamefont {Xu}, \citenamefont {Zhang}, \citenamefont {Mao}, \citenamefont {Wang}, \citenamefont {Xie},\ and\ \citenamefont {Zhang}}]{xu:metal:2020:curriculum}%
  \BibitemOpen
  \bibfield  {author} {\bibinfo {author} {\bibfnamefont {B.}~\bibnamefont {Xu}}, \bibinfo {author} {\bibfnamefont {L.}~\bibnamefont {Zhang}}, \bibinfo {author} {\bibfnamefont {Z.}~\bibnamefont {Mao}}, \bibinfo {author} {\bibfnamefont {Q.}~\bibnamefont {Wang}}, \bibinfo {author} {\bibfnamefont {H.}~\bibnamefont {Xie}},\ and\ \bibinfo {author} {\bibfnamefont {Y.}~\bibnamefont {Zhang}},\ }\bibfield  {title} {\bibinfo {title} {Curriculum learning for natural language understanding},\ }in\ \href {https://doi.org/10.18653/v1/2020.acl-main.542} {\emph {\bibinfo {booktitle} {Proc. 58th Annu. Meet. Assoc. Comput. Linguist.}}},\ \bibinfo {series and number} {ACL'20},\ \bibinfo {editor} {edited by\ \bibinfo {editor} {\bibfnamefont {D.}~\bibnamefont {Jurafsky}}, \bibinfo {editor} {\bibfnamefont {J.}~\bibnamefont {Chai}}, \bibinfo {editor} {\bibfnamefont {N.}~\bibnamefont {Schluter}},\ and\ \bibinfo {editor} {\bibfnamefont {J.}~\bibnamefont {Tetreault}}}\ (\bibinfo  {publisher} {Association for Computational
  Linguistics},\ \bibinfo {address} {Online},\ \bibinfo {year} {2020})\ pp.\ \bibinfo {pages} {6095--6104}\BibitemShut {NoStop}%
\bibitem [{\citenamefont {Narvekar}\ \emph {et~al.}(2020)\citenamefont {Narvekar}, \citenamefont {Peng}, \citenamefont {Leonetti}, \citenamefont {Sinapov}, \citenamefont {Taylor},\ and\ \citenamefont {Stone}}]{narvekar:2020:currsurvey}%
  \BibitemOpen
  \bibfield  {author} {\bibinfo {author} {\bibfnamefont {S.}~\bibnamefont {Narvekar}}, \bibinfo {author} {\bibfnamefont {B.}~\bibnamefont {Peng}}, \bibinfo {author} {\bibfnamefont {M.}~\bibnamefont {Leonetti}}, \bibinfo {author} {\bibfnamefont {J.}~\bibnamefont {Sinapov}}, \bibinfo {author} {\bibfnamefont {M.~E.}\ \bibnamefont {Taylor}},\ and\ \bibinfo {author} {\bibfnamefont {P.}~\bibnamefont {Stone}},\ }\bibfield  {title} {\bibinfo {title} {Curriculum learning for reinforcement learning domains: a framework and survey},\ }\bibfield  {journal} {\bibinfo  {journal} {J. Mach. Learn. Res.}\ }\textbf {\bibinfo {volume} {21}},\ \href {https://doi.org/10.5555/3455716.3455897} {10.5555/3455716.3455897} (\bibinfo {year} {2020})\BibitemShut {NoStop}%
\bibitem [{\citenamefont {Mari}\ \emph {et~al.}(2020)\citenamefont {Mari}, \citenamefont {Bromley}, \citenamefont {Izaac}, \citenamefont {Schuld},\ and\ \citenamefont {Killoran}}]{mari2020:quantum:transfer}%
  \BibitemOpen
  \bibfield  {author} {\bibinfo {author} {\bibfnamefont {A.}~\bibnamefont {Mari}}, \bibinfo {author} {\bibfnamefont {T.~R.}\ \bibnamefont {Bromley}}, \bibinfo {author} {\bibfnamefont {J.}~\bibnamefont {Izaac}}, \bibinfo {author} {\bibfnamefont {M.}~\bibnamefont {Schuld}},\ and\ \bibinfo {author} {\bibfnamefont {N.}~\bibnamefont {Killoran}},\ }\bibfield  {title} {\bibinfo {title} {Transfer learning in hybrid classical-quantum neural networks},\ }\href {https://doi.org/10.22331/q-2020-10-09-340} {\bibfield  {journal} {\bibinfo  {journal} {Quantum}\ }\textbf {\bibinfo {volume} {4}},\ \bibinfo {pages} {340} (\bibinfo {year} {2020})}\BibitemShut {NoStop}%
\bibitem [{\citenamefont {Kundu}\ and\ \citenamefont {Mangini}(2025)}]{kundu:2025:tensorrlqas}%
  \BibitemOpen
  \bibfield  {author} {\bibinfo {author} {\bibfnamefont {A.}~\bibnamefont {Kundu}}\ and\ \bibinfo {author} {\bibfnamefont {S.}~\bibnamefont {Mangini}},\ }\bibfield  {title} {\bibinfo {title} {Tensorrl-qas: Reinforcement learning with tensor networks for scalable quantum architecture search},\ }\bibfield  {journal} {\bibinfo  {journal} {arXiv}\ }\href {https://doi.org/10.48550/arxiv.2505.09371} {10.48550/arxiv.2505.09371} (\bibinfo {year} {2025})\BibitemShut {NoStop}%
\bibitem [{\citenamefont {Kanamori}\ \emph {et~al.}(2009)\citenamefont {Kanamori}, \citenamefont {Hido},\ and\ \citenamefont {Sugiyama}}]{kanamori:2009:jmlr}%
  \BibitemOpen
  \bibfield  {author} {\bibinfo {author} {\bibfnamefont {T.}~\bibnamefont {Kanamori}}, \bibinfo {author} {\bibfnamefont {S.}~\bibnamefont {Hido}},\ and\ \bibinfo {author} {\bibfnamefont {M.}~\bibnamefont {Sugiyama}},\ }\bibfield  {title} {\bibinfo {title} {A least-squares approach to direct importance estimation},\ }\href {http://jmlr.org/papers/v10/kanamori09a.html} {\bibfield  {journal} {\bibinfo  {journal} {J. Mach. Learn. Res.}\ }\textbf {\bibinfo {volume} {10}},\ \bibinfo {pages} {1391} (\bibinfo {year} {2009})}\BibitemShut {NoStop}%
\bibitem [{\citenamefont {Huang}\ \emph {et~al.}(2021)\citenamefont {Huang}, \citenamefont {Broughton}, \citenamefont {Mohseni}, \citenamefont {Babbush}, \citenamefont {Boixo}, \citenamefont {Neven},\ and\ \citenamefont {McClean}}]{huang:2021:power}%
  \BibitemOpen
  \bibfield  {author} {\bibinfo {author} {\bibfnamefont {H.-Y.}\ \bibnamefont {Huang}}, \bibinfo {author} {\bibfnamefont {M.}~\bibnamefont {Broughton}}, \bibinfo {author} {\bibfnamefont {M.}~\bibnamefont {Mohseni}}, \bibinfo {author} {\bibfnamefont {R.}~\bibnamefont {Babbush}}, \bibinfo {author} {\bibfnamefont {S.}~\bibnamefont {Boixo}}, \bibinfo {author} {\bibfnamefont {H.}~\bibnamefont {Neven}},\ and\ \bibinfo {author} {\bibfnamefont {J.~R.}\ \bibnamefont {McClean}},\ }\bibfield  {title} {\bibinfo {title} {Power of data in quantum machine learning},\ }\href {https://doi.org/10.1038/s41467-021-22539-9} {\bibfield  {journal} {\bibinfo  {journal} {Nat. Commun.}\ }\textbf {\bibinfo {volume} {12}},\ \bibinfo {pages} {2631} (\bibinfo {year} {2021})}\BibitemShut {NoStop}%
\bibitem [{\citenamefont {Huang}\ \emph {et~al.}(2022)\citenamefont {Huang}, \citenamefont {Kueng}, \citenamefont {Torlai}, \citenamefont {Albert},\ and\ \citenamefont {Preskill}}]{huang:2022:science}%
  \BibitemOpen
  \bibfield  {author} {\bibinfo {author} {\bibfnamefont {H.-Y.}\ \bibnamefont {Huang}}, \bibinfo {author} {\bibfnamefont {R.}~\bibnamefont {Kueng}}, \bibinfo {author} {\bibfnamefont {G.}~\bibnamefont {Torlai}}, \bibinfo {author} {\bibfnamefont {V.~V.}\ \bibnamefont {Albert}},\ and\ \bibinfo {author} {\bibfnamefont {J.}~\bibnamefont {Preskill}},\ }\bibfield  {title} {\bibinfo {title} {{Provably efficient machine learning for quantum many-body problems}},\ }\href {https://doi.org/10.1126/science.abk3333} {\bibfield  {journal} {\bibinfo  {journal} {Science}\ }\textbf {\bibinfo {volume} {377}},\ \bibinfo {pages} {eabk3333} (\bibinfo {year} {2022})}\BibitemShut {NoStop}%
\bibitem [{\citenamefont {Barkoutsos}\ \emph {et~al.}(2018)\citenamefont {Barkoutsos}, \citenamefont {Gonthier}, \citenamefont {Sokolov}, \citenamefont {Moll}, \citenamefont {Salis}, \citenamefont {Fuhrer}, \citenamefont {Ganzhorn}, \citenamefont {Egger}, \citenamefont {Troyer}, \citenamefont {Mezzacapo}, \citenamefont {Filipp},\ and\ \citenamefont {Tavernelli}}]{barkoutsos:2018:twolocal}%
  \BibitemOpen
  \bibfield  {author} {\bibinfo {author} {\bibfnamefont {P.~K.}\ \bibnamefont {Barkoutsos}}, \bibinfo {author} {\bibfnamefont {J.~F.}\ \bibnamefont {Gonthier}}, \bibinfo {author} {\bibfnamefont {I.}~\bibnamefont {Sokolov}}, \bibinfo {author} {\bibfnamefont {N.}~\bibnamefont {Moll}}, \bibinfo {author} {\bibfnamefont {G.}~\bibnamefont {Salis}}, \bibinfo {author} {\bibfnamefont {A.}~\bibnamefont {Fuhrer}}, \bibinfo {author} {\bibfnamefont {M.}~\bibnamefont {Ganzhorn}}, \bibinfo {author} {\bibfnamefont {D.~J.}\ \bibnamefont {Egger}}, \bibinfo {author} {\bibfnamefont {M.}~\bibnamefont {Troyer}}, \bibinfo {author} {\bibfnamefont {A.}~\bibnamefont {Mezzacapo}}, \bibinfo {author} {\bibfnamefont {S.}~\bibnamefont {Filipp}},\ and\ \bibinfo {author} {\bibfnamefont {I.}~\bibnamefont {Tavernelli}},\ }\bibfield  {title} {\bibinfo {title} {Quantum algorithms for electronic structure calculations: Particle-hole hamiltonian and optimized wave-function expansions},\ }\href {https://doi.org/10.1103/PhysRevA.98.022322}
  {\bibfield  {journal} {\bibinfo  {journal} {Phys. Rev. A}\ }\textbf {\bibinfo {volume} {98}},\ \bibinfo {pages} {022322} (\bibinfo {year} {2018})}\BibitemShut {NoStop}%
\bibitem [{\citenamefont {Gil-Fuster}\ \emph {et~al.}(2024{\natexlab{a}})\citenamefont {Gil-Fuster}, \citenamefont {Eisert},\ and\ \citenamefont {Bravo-Prieto}}]{fuster:2024:natcom}%
  \BibitemOpen
  \bibfield  {author} {\bibinfo {author} {\bibfnamefont {E.}~\bibnamefont {Gil-Fuster}}, \bibinfo {author} {\bibfnamefont {J.}~\bibnamefont {Eisert}},\ and\ \bibinfo {author} {\bibfnamefont {C.}~\bibnamefont {Bravo-Prieto}},\ }\bibfield  {title} {\bibinfo {title} {{Understanding quantum machine learning also requires rethinking generalization}},\ }\href {https://doi.org/10.1038/s41467-024-45882-z} {\bibfield  {journal} {\bibinfo  {journal} {Nat. Comm.}\ }\textbf {\bibinfo {volume} {15}},\ \bibinfo {pages} {2277} (\bibinfo {year} {2024}{\natexlab{a}})}\BibitemShut {NoStop}%
\bibitem [{\citenamefont {Recio-Armengol}\ \emph {et~al.}(2024)\citenamefont {Recio-Armengol}, \citenamefont {Schreiber}, \citenamefont {Eisert},\ and\ \citenamefont {Bravo-Prieto}}]{erik:2024:complexity}%
  \BibitemOpen
  \bibfield  {author} {\bibinfo {author} {\bibfnamefont {E.}~\bibnamefont {Recio-Armengol}}, \bibinfo {author} {\bibfnamefont {F.~J.}\ \bibnamefont {Schreiber}}, \bibinfo {author} {\bibfnamefont {J.}~\bibnamefont {Eisert}},\ and\ \bibinfo {author} {\bibfnamefont {C.}~\bibnamefont {Bravo-Prieto}},\ }\bibfield  {title} {\bibinfo {title} {Learning complexity gradually in quantum machine learning models},\ }\bibfield  {journal} {\bibinfo  {journal} {arXiv}\ }\href {https://doi.org/10.48550/arXiv.2411.11954} {10.48550/arXiv.2411.11954} (\bibinfo {year} {2024})\BibitemShut {NoStop}%
\bibitem [{\citenamefont {Cerezo}\ \emph {et~al.}(2023)\citenamefont {Cerezo}, \citenamefont {Larocca}, \citenamefont {García-Martín}, \citenamefont {Diaz}, \citenamefont {Braccia}, \citenamefont {Fontana}, \citenamefont {Rudolph}, \citenamefont {Bermejo}, \citenamefont {Ijaz}, \citenamefont {Thanasilp}, \citenamefont {Anschuetz},\ and\ \citenamefont {Holmes}}]{cerezo:2023:BP:sim}%
  \BibitemOpen
  \bibfield  {author} {\bibinfo {author} {\bibfnamefont {M.}~\bibnamefont {Cerezo}}, \bibinfo {author} {\bibfnamefont {M.}~\bibnamefont {Larocca}}, \bibinfo {author} {\bibfnamefont {D.}~\bibnamefont {García-Martín}}, \bibinfo {author} {\bibfnamefont {N.~L.}\ \bibnamefont {Diaz}}, \bibinfo {author} {\bibfnamefont {P.}~\bibnamefont {Braccia}}, \bibinfo {author} {\bibfnamefont {E.}~\bibnamefont {Fontana}}, \bibinfo {author} {\bibfnamefont {M.~S.}\ \bibnamefont {Rudolph}}, \bibinfo {author} {\bibfnamefont {P.}~\bibnamefont {Bermejo}}, \bibinfo {author} {\bibfnamefont {A.}~\bibnamefont {Ijaz}}, \bibinfo {author} {\bibfnamefont {S.}~\bibnamefont {Thanasilp}}, \bibinfo {author} {\bibfnamefont {E.~R.}\ \bibnamefont {Anschuetz}},\ and\ \bibinfo {author} {\bibfnamefont {Z.}~\bibnamefont {Holmes}},\ }\bibfield  {title} {\bibinfo {title} {{Does provable absence of barren plateaus imply classical simulability? Or, why we need to rethink variational quantum computing}},\ }\bibfield  {journal} {\bibinfo  {journal}
  {arXiv}\ }\href {https://doi.org/10.48550/arxiv.2312.09121} {10.48550/arxiv.2312.09121} (\bibinfo {year} {2023})\BibitemShut {NoStop}%
\bibitem [{\citenamefont {Gil-Fuster}\ \emph {et~al.}(2024{\natexlab{b}})\citenamefont {Gil-Fuster}, \citenamefont {Gyurik}, \citenamefont {Pérez-Salinas},\ and\ \citenamefont {Dunjko}}]{fuster:2024:VQA:dequan}%
  \BibitemOpen
  \bibfield  {author} {\bibinfo {author} {\bibfnamefont {E.}~\bibnamefont {Gil-Fuster}}, \bibinfo {author} {\bibfnamefont {C.}~\bibnamefont {Gyurik}}, \bibinfo {author} {\bibfnamefont {A.}~\bibnamefont {Pérez-Salinas}},\ and\ \bibinfo {author} {\bibfnamefont {V.}~\bibnamefont {Dunjko}},\ }\bibfield  {title} {\bibinfo {title} {On the relation between trainability and dequantization of variational quantum learning models},\ }\bibfield  {journal} {\bibinfo  {journal} {arXiv}\ }\href {https://doi.org/10.48550/arXiv.2406.07072} {10.48550/arXiv.2406.07072} (\bibinfo {year} {2024}{\natexlab{b}})\BibitemShut {NoStop}%
\bibitem [{\citenamefont {Mousavi~Kalan}\ \emph {et~al.}(2020)\citenamefont {Mousavi~Kalan}, \citenamefont {Fabian}, \citenamefont {Avestimehr},\ and\ \citenamefont {Soltanolkotabi}}]{kalan:2020:minimax}%
  \BibitemOpen
  \bibfield  {author} {\bibinfo {author} {\bibfnamefont {M.}~\bibnamefont {Mousavi~Kalan}}, \bibinfo {author} {\bibfnamefont {Z.}~\bibnamefont {Fabian}}, \bibinfo {author} {\bibfnamefont {S.}~\bibnamefont {Avestimehr}},\ and\ \bibinfo {author} {\bibfnamefont {M.}~\bibnamefont {Soltanolkotabi}},\ }\bibfield  {title} {\bibinfo {title} {Minimax lower bounds for transfer learning with linear and one-hidden layer neural networks},\ }in\ \href {https://proceedings.neurips.cc/paper/2020/hash/151d21647527d1079781ba6ae6571ffd-Abstract.html} {\emph {\bibinfo {booktitle} {Adv. Neural Inf. Process. Syst.}}},\ Vol.~\bibinfo {volume} {33},\ \bibinfo {editor} {edited by\ \bibinfo {editor} {\bibfnamefont {H.}~\bibnamefont {Larochelle}}, \bibinfo {editor} {\bibfnamefont {M.}~\bibnamefont {Ranzato}}, \bibinfo {editor} {\bibfnamefont {R.}~\bibnamefont {Hadsell}}, \bibinfo {editor} {\bibfnamefont {M.}~\bibnamefont {Balcan}},\ and\ \bibinfo {editor} {\bibfnamefont {H.}~\bibnamefont {Lin}}}\ (\bibinfo  {publisher} {Curran
  Associates, Inc.},\ \bibinfo {year} {2020})\ pp.\ \bibinfo {pages} {1959--1969}\BibitemShut {NoStop}%
\bibitem [{\citenamefont {mpmath~development team}(2023)}]{mpmath}%
  \BibitemOpen
  \bibfield  {author} {\bibinfo {author} {\bibfnamefont {T.}~\bibnamefont {mpmath~development team}},\ }\href@noop {} {\bibinfo {title} {mpmath: a {P}ython library for arbitrary-precision floating-point arithmetic (version 1.3.0)}} (\bibinfo {year} {2023}),\ \bibinfo {note} {{\tt http://mpmath.org/}}\BibitemShut {NoStop}%
\end{thebibliography}%

\end{document}